\documentclass[twocolumn,showpacs,amsmath,amssymb,prb]{revtex4-2}

\usepackage{
    graphicx,
    hyperref
}
\usepackage{color}
\graphicspath{{Figures/}}

\newcommand{\bl}{\begin{aligned}}
\newcommand{\el}{\end{aligned}}

\newcommand{\si}{\sigma}
\newcommand{\al}{\alpha}

\newcommand{\la}{\lambda}

\newcommand{\De}{\Delta}

\newcommand{\bOm}{{\bm\Omega}}

\newcommand{\bde}{{\bm\delta}}
\newcommand{\tbde}{\tilde{{\bm\delta}}}

\newcommand{\om}{i\omega_n}

\newcommand{\be}{\begin{equation}}
\newcommand{\ee}{\end{equation}}

\newcommand{\bea}{\begin{eqnarray}}
\newcommand{\eea}{\end{eqnarray}}
\newcommand{\bd}{\begin{displaymath}}
\newcommand{\ed}{\end{displaymath}}
\newcommand{\ba}{\begin{array}}
\newcommand{\ea}{\end{array}}
\newcommand{\bi}{\begin{itemize}}
\newcommand{\ei}{\end{itemize}}
\newcommand{\bc}{\begin{center}}
\newcommand{\ec}{\end{center}}
\newcommand{\bfl}{\begin{flushleft}}
\newcommand{\efl}{\end{flushleft}}
\newcommand{\bfr}{\begin{flushright}}
\newcommand{\efr}{\end{flushright}}

\newcommand{\hchi}{\hat{\chi}}

\newcommand{\hh}{\hat{h}}
\newcommand{\hH}{\hat{H}}

\newcommand{\tilm}{\tilde{m}}

\newcommand{\tgam}{\tilde{\gamma}}

\newcommand{\tilh}{\tilde{h}}

\newcommand{\bla}{\bar{\lambda}}
\newcommand{\bint}{\bar{I}}

\newcommand{\fs}{\frac{1}{2}}

\newcommand{\fhxa}{\frac{\sqrt{3}}{2}}
\newcommand{\fhxb}{\frac{\sqrt{3}}{3}}
\newcommand{\fhxc}{\frac{\sqrt{3}}{6}}

\newcommand{\blue}{\color{black}}
\newcommand{\bluex}{\color{black}}

\def\dg{^{\dagger}}
\def\bR{{\bf R}}
\def\bk{{\bf k}} \def\bK{{\bf K}}\def\bq{{\bf q}} 
  
\def\bQ{{\bf Q}} \def\bG{{\bf G}} \def\bJ{{\bf J}}
   \def\bv{{\bf v}}
 \def\bw{{\bf w}}
 \def\bd{{\bf d}}

\def\dg{\dagger}

\def\no{\nonumber \\}
\def\bra{\langle}
\def\ket{\rangle}
\def\={\!\!\!&=&\!\!\!}
\def\+{\!\!\!&&\!\!\!+~}
\def\-{\!\!\!&&\!\!\!-~}

\begin{document}
\date{\today}
\title{Topological paramagnetic excitons of localized f electrons on the honeycomb lattice}

\author{Alireza Akbari}
\author{Burkhard Schmidt}
\author{Peter Thalmeier}
\affiliation{Max Planck Institute for the Chemical Physics of Solids, 01187 Dresden, Germany}

\begin{abstract}
We investigate the dispersive paramagnetic excitons on the honeycomb lattice that originate from
the crystalline-electric field (CEF) split localized f-electron states in the {\it paramagnetic} state due to intersite exchange. We start with a symmetry analysis of possible Ising-type singlet-singlet and $xy$-type singlet-doublet models. The former supports only symmetric  intersite-exchange while the latter additionally allows for antisymmetric Dzyaloshinski-Moriya (DM) exchange  interactions. We calculate the closed expressions for  magnetic exciton dispersion using both response function formalism and bosonic Bogoliubov approach. We do this for the most general model that shows inversion symmetry breaking on the honeycomb lattice but also discuss  interesting special cases.  By calculating Berry curvatures and Chern numbers of paramagnetic excitons  we show that the $xy$ model supports nontrivial topological states in a wide range of parameters.
This leads to the existence of excitonic topological edge states with Dirac dispersion lying in the zone boundary gap without the presence of magnetic order. \end{abstract}


\maketitle

\section{Introduction}
\label{sec:introduction}

{\blue
Localized 4f and 5f electron states are organized in terms and multiplets according to Hund's rules. Since the
spin- orbit coupling is generally larger than crystalline electric field (CEF) potentials acting on the f-electrons
the total angular momentum of multiplets is a good quantum number. The perturbation of the CEF at the f-electrons site  
which originates from the surrounding ligands splits the ground state J-multiplet into a series of CEF mutliplets with degeneracies corresponding to the possible representations of the f- site symmetry. This is conveniently described within Steven's operator technique used in Refs. \onlinecite{hutchings:64,lea:62}  with an effective parametrized CEF Hamiltonian restricted to the lowest J- multiplet subspace (Appendix \ref{sec:appCEF}). The parameters may be formally expressed in terms of a point charge model (PCM) with screened ligand charges, however in practice they are usually determined from experiment.
The size of the splittings depend much on the material but are generally, at least for a subset of CEF multiplets  in the thermal range and lead to a large variety of physical effects \cite{jensen:91,fulde:72} for accessible temperatures. In particular the temperature dependence of the susceptibility over the whole range of CEF splitting allows to extract model sets of parameters for the CEF potential, which is, however, rarely unique.

 For intermetallic compounds the conduction electrons (c) have an effective on-site exchange interaction $J_{cf}$ with CEF states  (obtained from eliminating the cf- hybridisation and f-f Coulomb interaction \cite{coleman:15}). Firstly it will lead to a broadening of CEF excitations \cite{fulde:85} observable in inelastic neutron scattering (INS). If the CEF ground state is degenerate a low temperature Kondo effect results in coherent heavy fermion behaviour often accompanied by unconventional superconductivity \cite{thalmeier:05} . Furthermore the elimination of $J_{cf}$ leads to effective unretarded intersite exchange interactions of the RKKY tpe which for degenerate CEF  ground state  may cause magnetic order according to the Doniach phase diagram \cite{coleman:15}. But even in the paramagnetic state their presence entails the formation of collective magnetic exciton modes which can be viewed as propagating localized CEF excitations between the multiplets. These magnetic exciton modes have been found in numerous 4f compounds using INS \cite{jensen:91,fulde:85}. Determining the dispersion and intensity of magnetic excitons such experiments also allow  to identify suitable model Hamiltonians for the coupled CEF states by deriving multiplet splittings and inter-site exchange coupling models from comparison with theoretical results for the model \cite{buyers:75,houmann:79} . The latter are most conveniently obtained with the RPA- response function formalism of the dynamic magnetic susceptibility \cite{jensen:91} which we will also partly use in this work.
 
 Of particular interest are CEF systems with singlet nonmagnetic ground state as occurs for f-electron materials with integer J, e.g. Pr- and U- compounds (J=4). These cannot exhibit magnetic order of the conventional quasiclassical type by aligning pre-existing moments as in the case of degenerate magnetic CEF ground state. Rather the creation of local moments and their ordering appears simultaneously through quantum mechanical mixing of excited CEF states into 
the singlet ground state e.g. in two-singlet \cite{wang:68,wang:69} and three-singlet \cite{thalmeier:21} CEF level systems caused by intersite exchange.
This happens only when the latter is sufficiently large as expressed by a dimensionless control parameter $\xi$ (Sec.~\ref{sec:IsingRPA}) . If it is smaller than a critical value or negligible the compound stays paramagnetic \cite{birgeneau:71a}.

Such type of singlet ground state induced moment magnetism is preceded in the paramagnetic phase by a strong temperature dependence and a softening of a critical magnetic exciton mode to a varying degree at the ordering wave vector. This type of induced singlet-singlet magnetic order is found e.g. in Pr metal (under pressure) \cite{birgeneau:72,cooper:72,jensen:91}  and Pr compounds like PrSb \cite{mcwhan:79} and Pr$_3$Tl \cite{birgeneau:71,buyers:75}, PrCu$_2$\cite{kawarazaki:95}, PrNi \cite{savchenkov:19} and also in TbSb\cite{holden:74} and various U-compounds \cite{thalmeier:02,marino:23,marino:23a}. In the Pr systems the large hyperfine interaction with nuclear moments can also play an essential role in the ordering \cite{jensen:91}.

The mechanism of induced order is not restricted to dipolar magnetism, for example in YbRu$_2$Ge$_2$ the lowest $J=\frac{7}{2}$ Kramers doublets form a quasi quartet that supports induced quadrupolar order due to non-diagonal quadrupole matrix elements \cite{jeevan:06,takimoto:08} between them.\\

In these materials  it is frequently possible to restrict model calculations  to a reduced low energy level scheme consisting just of the singlet ground state and an excited multiplet (e.g. singlet or doublet) and ignoring the higher lying CEF states. Such simplified models will be also used in this work. They allow closed analytic solutions for the exciton bands and a detailed investigation how their structure and properties depend on the model parameters. 
Here we investigate the magnetic excitons for such simplified singlet ground state systems in the paramagnetic state where the f-electron sites are forming a 2D honeycomb lattice. It has a two-atom basis (A,B) (Fig.~\ref{fig:honeycomb}) each of them belonging to a trigonal Bravais lattice with site symmetry $C_{3v}$. The honeycomb lattice may be realized as a planar structure within a 3D lattice.
This structure is relevant for various f-electron compounds like Na$_2$PrO$_3$ \cite{daum:21}, TmNi$_3$Al$_9$\cite{Ge:22}  and recently a new class of promising 4f (RE =Tm,Ho) honeycomb materials BaRE$_2$(SiO$_4$)$_6$ has been discovered \cite{liu:23}. All compounds mentioned have integer total angular momentum $J$.  For concreteness we focus on $J=4$ realized in trivalent  Pr(4f$^2$) and possibly U(5f$^2$) magnetic ions but may also be applicable to trivalent Tb and Tm with $J=6$ and Ho with $J=8$. \\

We begin with an appropriate motivation why this is an interesting problem. 
It is already well known that in the  ferromagnetically (FM) or antiferromagnetically (AFM) {\it ordered} honeycomb lattice {\it magnon} bands may become topologically nontrivial and support magnonic edge modes within the gap of split 2D bulk magnon modes \cite{owerre:16a,owerre:16b,kim:16,kim:22,kondo:19,aguilera:20,kondo:21}. 
This well developed subject is reviewed in Refs.~\onlinecite{kondo:15,bonbien:21,mcclarty:22,zhuo:23,yu:23}.
The nontrivial topology in 2D is characterized by a nonzero Chern number of the bulk bands which is the integral over the Berry curvature obtained from the magnon bands and their eigenstates. The gap opening between the two magnon bands (due to sublattice structure) is a prerequesite for nonvanishing Chern number. It can only be achieved if an antisymmetric Dzyaloshinskii-Moriya (DM) spin exchange term between nearest neighbors  is included. Any symmetric exchange (between first neighbors on A,B or between further neighbors) will preserve the degeneracy of magnon bands at zone boundary points of the trigonal Brillouin zone (BZ) leading to Chern number zero. The DM interaction is allowed because the centers of n.n. A-B are not inversion centers of the lattice, only the centers of hexagons and n.n.n. bonds (Fig.~\ref{fig:honeycomb}). The DM interaction thus enables nonzero Chern number and consequently (nondegenerate) magnon edge states inside
the bulk gap. They can carry a transverse heat current thus leading to a topological thermal magnon Hall and Nernst effect discussed in theoretical investigations, e.g., Refs \onlinecite{owerre:16b,cheng:16} and found experimentally in a similar kagome lattice FM  \cite{madhorgaria:23}.  \\

In this work we will study the {\it paramagnetic excitons} on the honeycomb lattice with nonmagnetic singlet ground
state f-electrons on the C$_{3v}$ sites having in mind the potentially interesting topological properties in analogy to the mangonic case. The aim of the present  work is twofold:}

 Firstly we want to give a complete theory of magnetic excitons in the paramagnetic state for CEF split f-electrons on the honeycomb lattice comprising two trigonal sublattices A,B and $C_{3v}$ site symmetry based on the reduced level schemes.  We focus on two representative cases for $C_{3v}$ CEF states: An Ising-type singlet-singlet system and an xy-type singlet-doublet level scheme. Thereby we make the most general assumption that inversion symmetry is broken leading to inequivalent CEF splitting and interaction parameters for sublattices A,B.
{\blue The aim of this part is to give a solid theoretical foundation for  inelastic neutron scattering (INS) experiments  on singet ground state honeycomb f-electron paramagnets. We will derive general model
expressions for dispersions and intensities that may be used to analyze such experiments provided one a restriction to one excited singlet or doublet can be justified, as is frequently the case in Pr- and U- compounds.}

{\bluex Characteristically the magnetic excitons appear already in the paramagnetic phase of singlet ground state systems as opposed to magnons which are seen only in the ordered phase as collective excitations of the order parameter resulting from a degenerate magnetic ground state and thus they are clearly separate types of magnetic excitations.  In an INS experiment both magnetic excitons and magnons can be distinguished in a standard way from phonon excitations of the underlying lattice by following their intensity  as function of total momentum transfer $\tilde{\bk}$ (including the reciprocal lattice vector). In the former the intensity decreases with $\tilde{\bk}$ due to the magnetic f-electron form factors while in the latter it increases quadratically with $\tilde{\bk}$ \cite{fulde:85}. The magnetic excitons considered here bear some formal similarity to the zero-field dispersive triplon excitations of spin dimer compounds between singlet and excited triplet state of the dimer \cite{matsumoto:04}. The dispersion is  caused by inter-dimer exchange smaller than the dimer singlet-triplet gap. However, such suitably sized dimerization is not relevant in any of the abovementioned compounds and also not in the honeycomb lattice discussed here with only equidistant f-electron sites. }

The Ising-type model is convenient for demonstrating the two techniques of calculating the magnetic exciton modes, namely the RPA response function and bosonic Bogoliubov quasiparticle techniques. We will show that indeed they give equivalent results. Applied to the Ising case we calculate the dispersion and intensity of the two modes symmetrically split by the inter-sublattice interactions and an additional contribution  resulting from the intra-sublattice terms. For equivalent sublattices the modes will be degenerate at specific zone boundary points $\bK_\pm$ and we demonstrate how they will be split when inversion symmetry breaking occurs.

Using the same techniques we investigate the richer singlet-doublet xy-type model. Because of nonzero diagonal matrix elements for both $J_x,J_y$ total angular momentum components an asymmetric DM interaction is possible for the intra-sublattice exchange. Due to the doublet degeneracy four magnetic exciton modes exist in principle. For equivalent sublattices they consist of a pair of twofold degenerate modes which can develop a gap at the $\bK_\pm$ zone boundary due to the presence of the DM interaction. A further splitting into four modes occurs when the sublattices become inequivalent. 
This theory is sufficiently general to be used for modeling INS experiments for all possible singlet-singlet and singlet-doublet CEF systems on compounds with f-electrons located on the honeycomb lattice.\\

Secondly we show that in the xy-type model the DM term may lead to interesting nontrivial topology of the magnetic exciton bands. We stress that this happens in the {\it paramagnetic} state of f electrons on the honeycomb lattice. It is our primary intention to demonstrate that magnetic order is not a prerequisite for the existence of topological magnetic excitations and corresponding edge modes. For this purpose we investigate the behaviour of Berry curvature and associated Chern numbers of paramagnetic exciton bands and discuss their model parameter dependence. We show that as function of the size of inversion symmetry breaking transitions from zero to integer Chern numbers is possible. In the latter case we also derive the existence of the boundary magnetic exciton modes in a continuum approximation around the Dirac points $\bK_\pm$. Finally we discuss, that in contrast to topological magnons in a FM the paramagnetic topological magnetic excitons do not lead to a thermal Hall effect as is indeed required by the absence of time reversal symmetry breaking.

In Sec.~\ref{sec:CEF} we give a brief introduction to f-electron CEF states in less common $C_{3v}$ symmetry with details relegated to  Appendix \ref{sec:appCEF}. Then Sec.~\ref{sec:Isingmod} discusses the Ising-type models in various techniques and the principle of induced magnetic order. In Sec.~\ref{sec:xymod} the xy-type model, its characteristic four dispersion branches and their topological properties including edge modes are investigated. Sec.~\ref{sec:discussion} discusses some numerical results and finally Sec.~\ref{sec:conclusion} gives the summary and conclusion.

\section{CEF states on the honeycomb lattice, singlet-singlet and singlet doublet models}
\label{sec:CEF}

\begin{figure}
\includegraphics[width=1.0\linewidth]{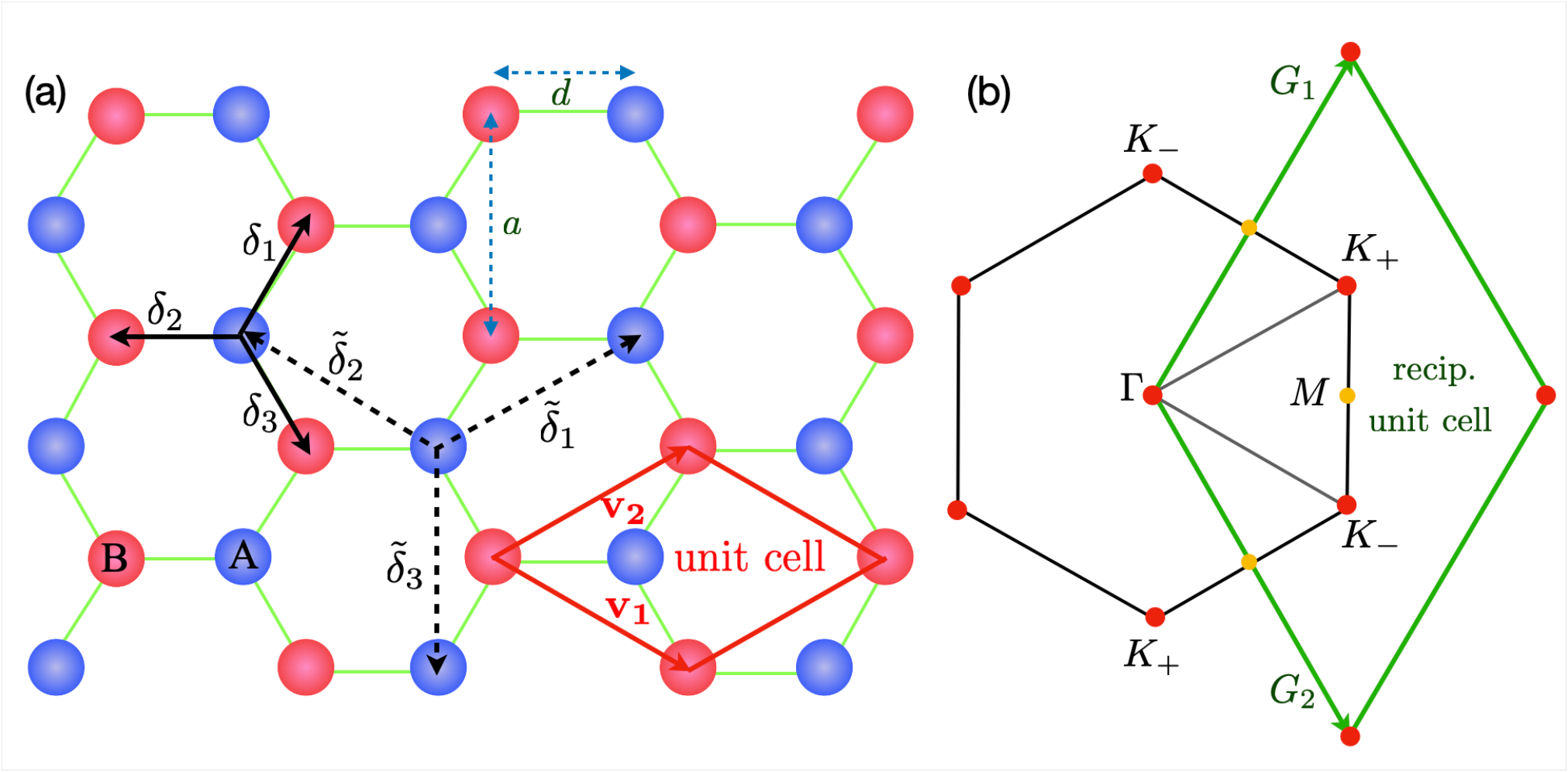}
\caption{(a) Honeycomb lattice structure (triangular sublattices $\si=A,B$) with unit cell (primitive lattice vectors $\bv_{1,2}$, $1^\text{st}$ neighbor vectors $\bde_i (i=1-3, z=3)$ and 
$2^\text{nd}$ neighbor vectors $\pm\tbde_i$ $(i=1-3, z=6)$ indicated. The corresponding symmetric exchange constants are $I_\si$ and $I^\si _2$, respectively. Lattice constant denoted by $a$ and $d=a/\sqrt{3}$ is the $1^\text{st}$ neighbor distance.
The DM exchange couplings between $2^\text{nd}$ neighbors at $\tbde_i,-\tbde_i$ are $-D^\si_J,D^\si_J$ respectively.
The distance between the zigzag chains (e.g. along $y$ direction is given by $x_0=\frac{\sqrt{3}}{2}$.
(b) Reciprocal lattice with primitive unit cell and associated vectors $\bG_1, \bG_2$. The inequivalent zone boundary valleys are indicated by $\bK_+,\bK_-$. The expressions for the vectors in direct and reciprocal space are given in Appendix \ref{sec:appgeometry}.}
\label{fig:honeycomb}
\end{figure}

The point group symmetry for the sites on the 2D honeycomb lattice with two basis atoms (A,B) is $C_{3v}$, composed of threefold rotations and reflections on perpendicular planes $120^\circ$ apart (Fig.\ref{fig:honeycomb}). The A,B sublattice sites have no inversion symmetry in $C_{3v}$. 
The honeycomb space group $P6/mcc$,  however, contains the inversion with centers given by the midpoint of bonds and the center of hexagons. The point group symmetry leads to a CEF potential (restricted to the lowest $J$-multiplet) given as a sum of Stevens operators $O_n^m(\bJ)$ $(m\leq n \leq 6)$ (see detailed analysis in Appendix  \ref{sec:appCEF}).

In this work we are interested exclusively in f-electron shells with integer $J$ to have the possibility
of a nonmagnetic singlet  CEF ground state $|0\rangle$ with $\langle 0| \bJ |0\rangle =0$. Among
the trivalent rare earth (RE) ions this is possible for $J=4$ (Pr),  $J=6$ (Tb,Tm) and $J=8$ (Ho). We will  restrict 
to the simplest case of $J=4$. The complete characterization of CEF energies and states in $C_{3v}$ symmetry 
is given in Appendix \ref{sec:appCEF}. In this group the $J=4$ space decomposes into irreducible representations 
$2\Gamma_1\oplus\Gamma_2\oplus 3\Gamma_3$, i.e. three singlets $(\Gamma_1^{a,b},\Gamma_2)$ and three doublets $(\Gamma_3^{a,b,c})$ which are linear combinations  of free ion states $|J,M\rangle$ $(|M|\leq J)$. The two $\Gamma_1^{a,b}$ singlets are characterized by one $(\theta)$ and the three $\Gamma^{a,b,c}_3$ doublets by generally three $(\chi,\phi,\alpha)$  mixing angles determined by the set of CEF parameters $B_m^n$ in Eq.~(\ref{eq:CEFpot}) while the unique $\Gamma_2$ is fully determined by $C_{3v}$ symmetry. Explicitly the full orthonormal CEF state basis is given in Appendix \ref{sec:appCEF}. Here we list only the singlets and one representative doublet $\Gamma_3^a$ necessary for the following analysis:
\be
\bl
    |\Gamma_{1a}\ket
    &=
    \hphantom{-}\cos\theta|4,0\ket+\frac1{\sqrt2}\sin\theta(|4,3\ket -|4,-3\ket),
    \\
    |\Gamma_{1b}\ket
    &=
    -\sin\theta|4,0\ket+\frac1{\sqrt2}\cos\theta(|4,3\ket-|4,-3\ket),
    \\
    |\Gamma_2\ket
    &=
    \frac1{\sqrt2}(|4,3\ket+|4,-3\ket),
    \\
    |\Gamma^\pm_{3a}\ket
    &=
    \sin\chi(\cos\phi|4,\pm4\ket+\sin\phi|4,\mp2\ket)
    \\
    &\phantom{=}
    \pm\cos\chi|4,\pm1\ket.
\el
\ee
The CEF energies $E_\Gamma$ of these eigenstates are complicated combinations of the $B_n^m$ (Appendix \ref{sec:appCEF}). Because there are six independent parameters and six irreducible representations the energy levels can in principle take any ordering.

 For investigating the magnetic exciton modes it is important to calculate the dipolar matrix elements between the CEF states. 
  The $J_\al (\al = x, y, z)$ operators connect states with $M' = M, M \pm 1$.
  Here we restrict to two important cases discussed in detail in the following: The singlet-singlet $\Gamma_{1a,b}$-$\Gamma_2$ subspaces and the singlet-doublet $\Gamma_2$-$\Gamma_{3a}$ subspaces. Their dipolar matrix elements are given by
 \be
 \bl
 \bra \Gamma_2|J_z|\Gamma_1\ket =& m; 
  \\ 
 \bra \Gamma_2|J_x|\Gamma_3^\pm\ket =&\tilde{m}/\sqrt{2};
 \quad 
 \bra \Gamma_2|J_y|\Gamma_3^\pm\ket =
 \pm{\rm i}\tilde{m}/\sqrt{2},
  \label{eq:matel}
  \el
 \ee
where we defined $m=3\sin\theta$ or  $m=3\cos\theta$ for $\Gamma_{1a,b}$ singlets, respectively and  $\tilde{m}=(1/\sqrt2)\sin\chi(\sqrt{7}\sin\phi+2\cos\phi))$ for $\Gamma_{3a}$. {\blue The matrix elements of $J_x,J_y$ between the $\Gamma_{1a,b}$-$\Gamma_2$ subspaces  vanish as well as those within $\Gamma_{3a}$ doublet subspace.}  Therefore the singlet-singlet $\Gamma_1$-$\Gamma_2$  model is of the Ising type while the singlet-doublet model $\Gamma_2$-$\Gamma_3$ is of the xy type {\blue for the inelastic CEF excitations}. The latter would also be realized in a $\Gamma_1$-$\Gamma_3$ type model. These selection rules follow also directly from the group multiplication table~\cite{koster:63} of $C_{3v}$ considering the fact that $J_z$ transforms like $\Gamma_2$ and $(J_x,J_y)$ transform like $\Gamma_3$. {\blue We note that nondiagonal quadrupolar matrix elements between ground- and excited state are only allowed for the xy-type model. Quadrupolar intersite interaction terms will  not be included here as they contribute only indirectly to the dipolar dynamic repsonse functions of INS in zero field \cite{portnichenko:20}.}

To devise suitably general models for both cases in the following sections we start from two basic observations on the honeycomb structure: Firstly the center of $2^\text{nd}$ neighbor bonds (A-A, B-B) is not an inversion center. Therefore in addition to symmetric exchange asymmetric Dzyaloshinski-Moriya (DM) exchange between $2^\text{nd}$ neighbors (dashed lines in Fig.~\ref{fig:honeycomb}) may be present. {\blue Secondly although the bond center of $1^\text{st}$ neighbours are inversion centers meaning that A,B sublattices are equivalent, this can be removed 
when the 2D honeycomb lattice is placed into a 3D crystal where the chemical environment of the basis atoms A, B between the honeycomb layers may be different.  This could be achieved by sandwiching the f-electron honeycomb layer between nonmagnetic honeycomb layers with different chemical occupations of A,B  known, e.g., from unconventional honeycomb superconductors  \cite{kudo:18}. Using  such 3D layered structure with local inversion symmetry breaking on the f- honeycomb sites} their CEF potentials (multiplet splittings) and interactions on the A, B sublattices may also be generally different. This possibility should be incorporated in both models. It means that inversion symmetry with respect to center of  $1^\text{st}$ neighbor A-B bonds and hexagon centers is also broken. {\blue We stress that such full 2D inversion symmetry breaking in honeycomb models has already  been proposed and  investigated before for the FM {\it ordered} honeycomb lattice~\cite{kim:22}}.

\section{The singlet-singlet Ising-type model}
\label{sec:Isingmod}

First we address the more simple and instructive case of the singlet-singlet CEF model.
Our calculations of exciton modes will be based on RPA response function theory as
well as Bogoliubov transformation approach. The former can also be applied at finite
temperatures while the latter allows to address topological properties of the modes
due to a bosonic representation used for the local CEF excitations.

For concreteness we assume $\Gamma_2$ to be the ground state and one of the $\Gamma_{1a,b}$  the excited state, the inverted scheme leads to identical results. Furthermore we do not distinguish between a, b representations and denote by $m=m_a,m_b$ any of the two matrix elements between ground and excited state. The singlet-singlet CEF Hamiltionian is then given by
\be
\bl
H=\sum_{\Gamma\sigma i}E_\Gamma^\sigma|\Gamma_{\sigma i} \ket\bra\Gamma_{\sigma i}|
-I\sum_{\bra ij\ket}J^z_{iA}J^z_{jB}
-\sum_{\bra\bra ij\ket\ket\sigma}I_2^\si J^z_{i\sigma}J^z_{j\sigma}
.
\label{eq:HIsing}
\el
\ee
Here $\sigma=A,B$ denotes the two sublattices and $i,j$ the $1^\text{st}$ neighbor lattice sites on each of them and $\Gamma=\Gamma_2,\Gamma_1$ the two singlet states. In the first term the CEF energies $E_{\Gamma\sigma}$ (and the $\Gamma_{1a,b}$ excited states) may depend on the sublattice A, B and similar for the exchange terms.  We fix $E^\sigma_{\Gamma_2}=0$ on each and denote the relative exited state energy by $\Delta_{\sigma}=E_{\Gamma_1\sigma}$ (we suppress a,b index of both possible $\Gamma_{1a,b}$ representations from now on). The second and third terms describe the symmetric exchange {\em between\/} A, B sublattices ($1^\text{st}$ neighbors) and  {\em within\/} A and B sublattices ($2^\text{nd}$ neighbors), respectively. {\blue Having in mind intermetallic f- electron compounds the effective intersite exchange terms  may be generated by the virtual exchange of e.g. 5d,6s- conduction electron-hole excitations \cite{hanzawa:19,yamada:19}.} Note that in the above model only $J_z$ has nonzero matrix elements (Eq.~(\ref{eq:matel})). Therefore it is of the Ising-type and in particular no DM exchange is supported because this needs at least two components of $\bJ$ to have nonzero matrix elements (Sec.~\ref{sec:xymod}).

\subsection{Response functions and magnetic exciton modes}
\label{sec:IsingRPA}

The interaction terms in Hamiltonian of Eq.~(\ref{eq:HIsing}) allow the $\Gamma_2 \leftrightarrow \Gamma_1$ excitations of the paramagnetic state to propagate from site to site and thus acquire a dispersion. They are commonly designated 'magnetic excitons' to distinguish them from magnons which
require a {\it magnetically ordered} ground state with broken time reversal symmetry.
The most convenient way to obtain the dispersion of magnetic excitons is the calculation of the dynamic magnetic susceptibility $\hat{\chi}(\bq,\om)$ in RPA. It is given by the $2\times 2$ sublattice-space matrix
\be
\bl
\hat{\chi}(\bk,\om)
&=
[1-\hat{I}(\bk)\hat{u}({\om})]^{-1}\hat{u}({\om})
,
\el
\ee
where
\be
\bl
 \hat{u}({\om})&=
\left(
 \begin{array}{cc}
 u_A(\om)& 0 \\
 0& u_B(\om)
\end{array}
\right),
\el
\ee
and
\be
\bl
\hat{I}({\bk})
&=
\left(
 \begin{array}{cc}
 z_2I^A_2\gamma_2(\bk)& zI\gamma(\bk) \\
 zI\gamma^*(\bk)&  z_2I^B_2\gamma_2(\bk)
\end{array}
\right)
\label{eq:RPA}
\el
\ee
%
 are  the single ion susceptibility and exchange matrices, respectively. In the latter $z=3$ and $z_2=6$ are first and second neighbor coordination number and $\gamma(\bk) ,\gamma_2(\bk)$ the corresponding structure functions of the honeycomb lattice (Eq.~(\ref{eq:appstruc2})).{\blue We note that the above exchange model for the 2D honeycomb can easily be generalized to a 3D stacked arrangement by introducing additional inter-layer exchange contstants and appropriately modified 3D structure functions. The exchange functions Eq.(\ref{eq:interaction}) for the xy-type model may be generalized in a similar fashion.} 
 
 Furthermore in the singlet-singlet model we have $(\sigma=A,B)$:
\be
u_{\sigma}(\om)=\frac{2m^2_\sigma\De_\sigma P_\si(T)}
{\De^2_\sigma-(\om)^2}.
\ee
The temperature dependent factor $P_\sigma(T)= \tanh\frac{\De_\sigma}{2T}$ in the numerator is equal to the difference of thermal occupations of ground and excited singlet state and $\Delta_\sigma$ and $m_\sigma$ are the (generally different) singlet-singlet splitting and matrix elements. The magnetic exciton bands (there are two $(\kappa=\pm)$ due two the A,B sublattices) are then obtained as the collective modes, i.e. the singularities of the dynamic susceptibility as determined by $det \hat{\chi}(\bk,\om)=0$. Solving this equation a closed expression for the magnetic exciton dispersions $\omega_\kappa(\bk)$ may be evaluated:
\be\bl
\omega_\pm^2(\bk)
=&
\fs[\omega_A^2(\bk)+\omega_B^2(\bk)]
\pm
\Big[
\frac{1}{4}(\omega_A^2(\bk)-\omega_B^2(\bk))^2
+
\\&
4m_A^2m_B^2\De_A\De_BP_AP_B|I_N(\bk)|^2
\Big]^\fs;
\\
\omega_\sigma^2(\bk)
=&\De_\si[\De_\si-2m_\si^2P_\si I^\si_D(\bk)]
.
\el
\ee
Here we use the abbreviations $I^\si_D(\bk)= (z_2I^\si_2)\gamma_2(\bk)$ and $I_N(\bk)= (zI)\gamma(\bk)$ for diagonal (D)
and nondiagonal (N) intra- and inter-sublattice exchange in Eq.~(\ref{eq:RPA}), respectively.
Furthermore the  $\omega_{A,B}(\bk)$ may be interpreted as the separate mode dispersions on $\sigma=$A,B sublattices when the nearest neighbor inter-sublattice
coupling $I_N(\bk)$ is set to zero. Explicitly this formula may also be written as 
\be
\bl
&\omega^2_\pm(\bk)=
\\
&\fs
(
\De_A^2+\De_B^2)
- 
[m_A^2\De_AP_AI^A_D(\bk)+m_B^2\De_BP_BI^B_D(\bk)
]
\pm
\\
&
\Bigl\{\bigl[\fs(\De_A^2-\De_B^2)-(m_A^2\De_AP_AI^A_D(\bk)-m_B^2\De_BP_BI^B_D(\bk))\bigr]^2
\\
&+4m_A^2m_B^2\De_A\De_BP_AP_B|I_N(\bk)|^2\Bigr\}^\fs.
\label{eq:RPAdisp0}
\el
\ee
%
%

\begin{figure}
\includegraphics[width=0.8\linewidth]{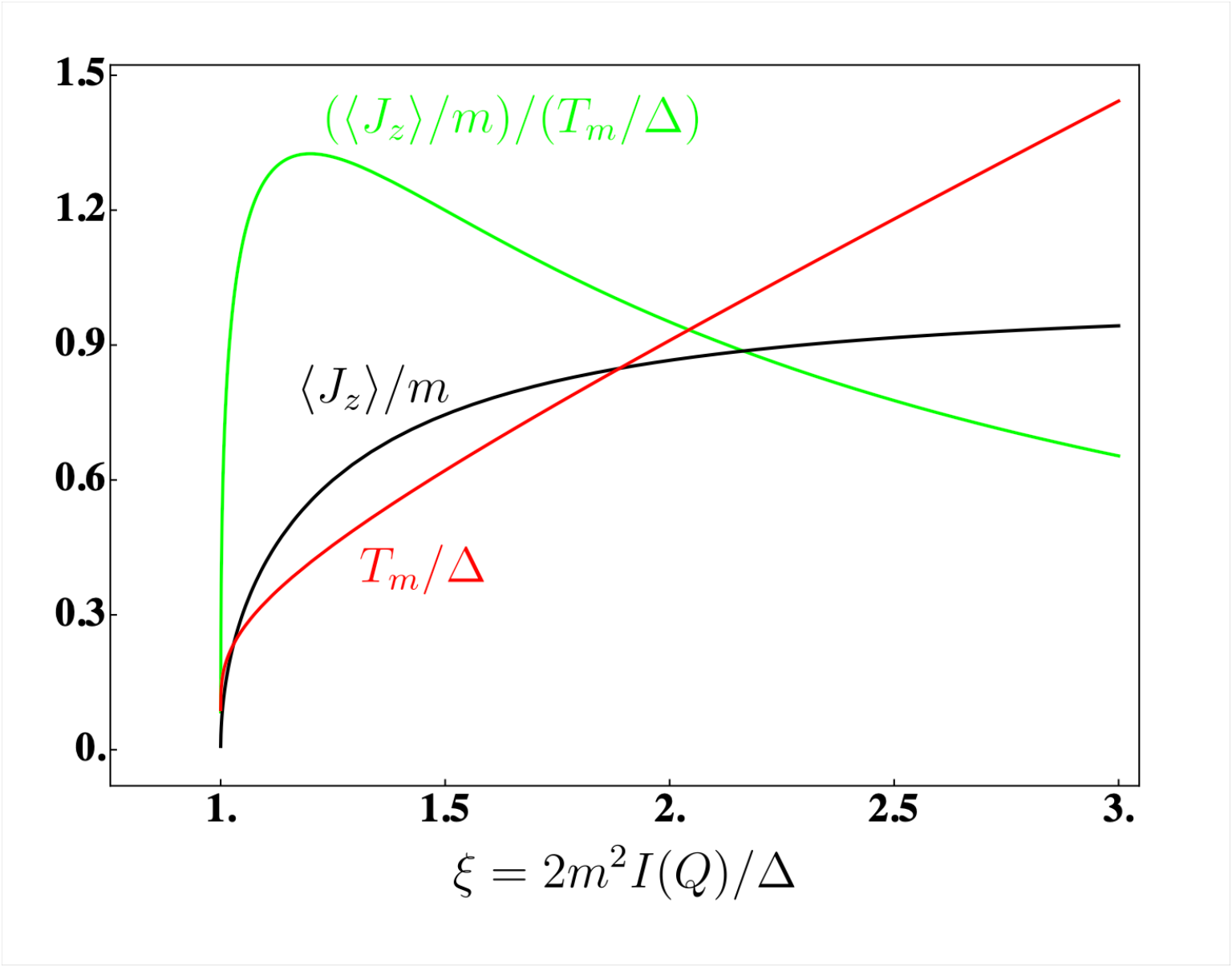}
\caption{Ising-type model induced order characteristics signified by control-parameter $\xi$-dependence of ground state moment $\langle J_z\rangle$ (normalized to $m$), magnetic ordering temperature T$_m$ (normalized to CEF splitting $\Delta$) and their ratio, see also Ref.~\onlinecite{marino:23a}.}
\label{fig:singsing}
\end{figure}

For numerical calculations it is convenient to use three model parameters (dimension energy) $v_s=(m_Am_BI)$ and $v_2^\si=(m_\si^2I_2^\si)$ and likewise  $|\bar{I}_N(\bk)|=m_Am_B| I_N(\bk)|=(zv_s)\gamma(\bk)$ and $\bar{I}_D^\si(\bk)=m_\si^2I_D^\si(\bk)=(z_2v_2^\si)\gamma_2(\bk)$
(see also Appendix \ref{sec:parameters}). At low temperatures $T/\De_\si\ll1$
we may replace $P_\sigma(T) \rightarrow 1$. The dispersion simplifies further if the intra-sublattice exchange $I^\si_D(\bk)$ 
is absent. Then we get
\be
\bl
&
\omega^2_\pm(\bk)
=
\fs(\De_A^2+\De_B^2)
\pm
\\
&
\hspace{0.4cm}
\bigl[\frac{1}{4}(\De_A^2-\De_B^2)^2+
4m_A^2m_B^2\De_A\De_BP_AP_B|I_N(\bk)|^2\bigr]^\fs
.
\label{eq:RPAdisp1}
\el
\ee
On the other hand if both $1^\text{st}$ and $2^\text{nd}$ neighbour exchange are kept but the two sublattice sites are assumed equivalent with $\De_A=\De_B=\De$ and likewise  $I_D^A=I_D^B=I_D$ Eq.(\ref{eq:RPAdisp0}) reduces to
\be
\bl
\omega_\pm^2(\bk)=\De\bigl[\De-2m^2(I_D(\bk)\mp|I_N(\bk)|)\tanh\frac{\De}{2T}\bigr]
.
\label{eq:RPAdisp2}
\el
\ee
Here the mode splitting of  $\omega_\kappa(\bk)$ can be seen to be directly associated with the inter-sublattice coupling. The splitting vanishes at the \bK$_\pm$ zone boundary points in this special case. In the general case described by Eq.~(\ref{eq:RPAdisp0}) the criterion for opening a gap at \bK$_\pm$ may be identified as i) for $\Delta_A\neq\Delta_B$ the gap is always present and ii) for $\Delta_A=\Delta_B$ one then must have $I^A_2\neq I^B_2$ for the intra-sublattice exchange. {\blue Furthermore we can see from the above special case that the band width of magnetic excitons is 
controlled by the size and  \bk-dependence of exchange interactions, increasing with their strength. It is frequently comparable to the CEF splitting $\Delta$ \cite{rainford:71,savchenkov:19}.}\\

\begin{figure}
\includegraphics[width=1\linewidth]{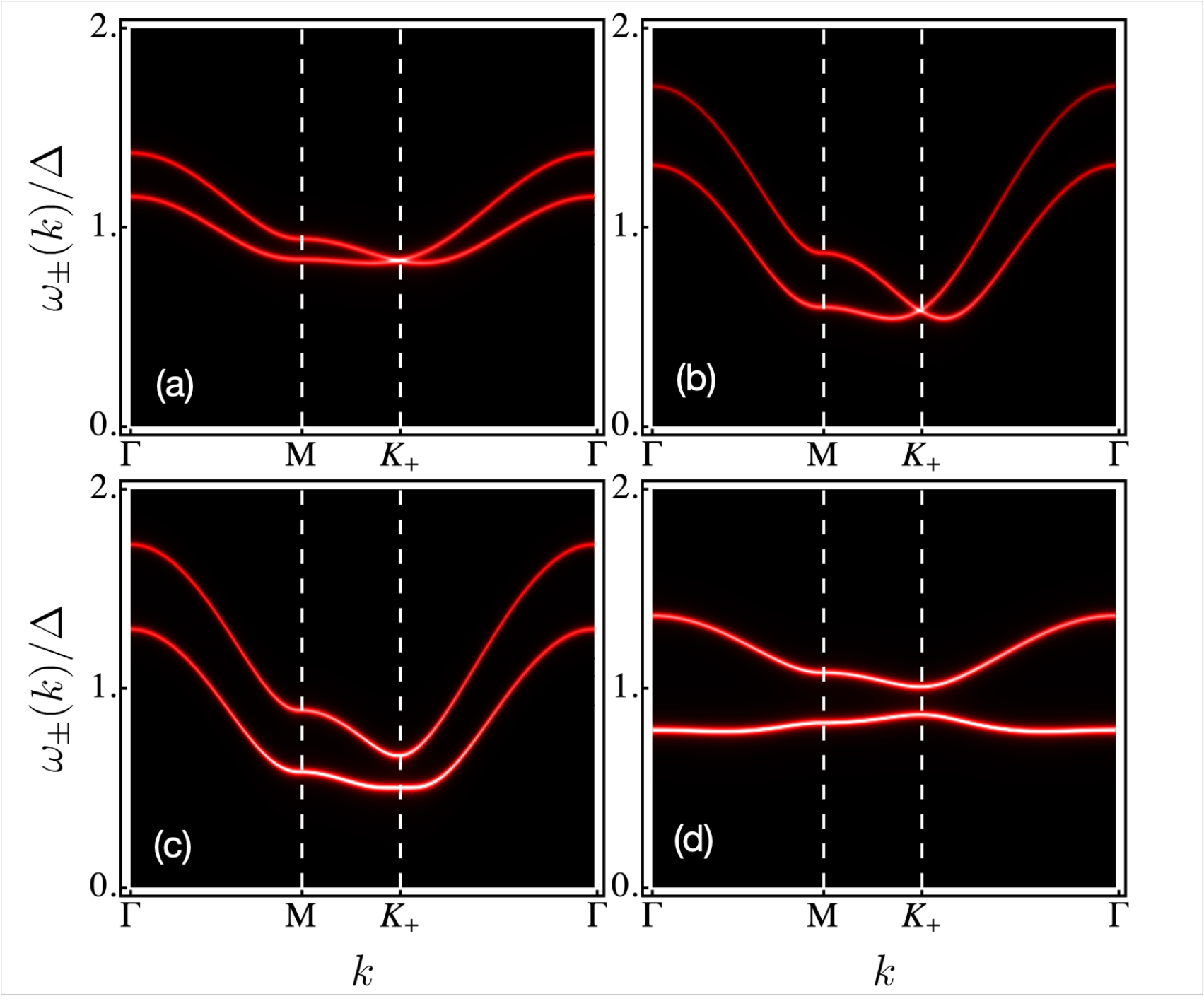}
\caption{Typical cases of the Ising model magnetic exciton dispersions. (a) high temperature case $T=1.0$ with equal $\Delta_{A,B}=1$,
 $v^{A,B}_2=-0.11$ and $v_s=-0.10$ shows moderate dispersion. (b) Same parameters but low temperature case exhibits large dispersion due 
 to increase thermal population difference of $\Gamma_2, \Gamma_3$ levels. Because of A,B equivalent interaction constants $K_+$ (and also $K_-$) is a Dirac point with degenerate and linearly dispersive exciton modes. The splitting of modes for all other \bk-values is due to inter-sublattice interaction $I_N(\bk)\sim v_s$. (c) $T=0.1$ case now with distinct $\Delta_{A,B}=\Delta(1\pm\epsilon)$ where $\Delta=1.$ and $\epsilon=0.07$ and other constants as in (a,b). Now the degeneracy at $K_\pm$ is removed. This case shows incipient soft mode behaviour around $K_+$ indicating closeness to commensurate spiral order. (c) Same case but small $v_2^{A,B}=-0.02$ which reduces the overall dispersion.
}
\label{fig:isingdisp}
\end{figure}

{\blue Eventually if the interactions become strong enough} the lower mode, e.g. $\omega_-(\bk)$ may become soft at specific, generally incommensurate wave vector \bk=\bQ~and this heralds a spontaneous induced magnetic order with modulation wave vector \bQ~of the singlet-singlet system although both CEF singlets are nonmagnetic with $\bra \Gamma_\alpha | J_z| \Gamma_\alpha\ket=0$ $(\alpha=1,2)$. In the above equivalent sublattice case this occurs when the control parameter
\be
\xi=\frac{2m^2I(\bQ)}{\De} > 1,
\ee
where $I(\bQ)=I_D(\bQ +|I_N(\bQ)|$ is the total exchange Fourier transform. For $\xi >1$ the transition temperature $T_m$ to the induced moment phase and the size of the induced moment $M_\bQ=\langle J_z\rangle$ (in units of $\mu_B$) along z are given by \cite{thalmeier:02}
\be
\bl
&
T_m\simeq\frac{\Delta}{2\tanh^{-1}\bigl(\frac{1}{\xi}\bigr)}\simeq\frac{\Delta}{|ln\xi'|};
\\
&
M_\bQ/m=\frac{1}{\xi}(\xi^2-1)^\fs\simeq \bigl(2\xi'\bigr)^\fs,
\el
\ee
where the approximate expressions hold close to the critical control parameter i.e. $\xi\simeq 1+\xi'$ with $\xi'\ll 1$. Both quantities increase with infinite slope above $\xi=1$  (Fig.~\ref{fig:singsing}). This Ising  type 2-singlet induced moment system has also been generalized for the frequently occurring three-singlet model in low symmetry 4f and 5f materials \cite{thalmeier:21}. In the present case when the incipient soft mode $(\xi <1)$ appears at $\bQ=\bK_\pm$ zone boundary positions as is the case in Fig.~\ref{fig:isingdisp} the magnetic order for critical $\xi=1$  would correspond to a $120^\circ$ commensurate spiral structure on each triangular sublattice A,B coupled ferro- or antiferromagnetically depending on the sign of intersublattice coupling $I$ in Eq.~(\ref{eq:HIsing}).  \\

In this work, however, we restrict to the investigation to the {\it paramagnetic} phase for both CEF models. 
In the response function formalism it is also straightforward to calculate the momentum and temperature dependence of the intensity of paramagnetic exciton modes 
that are essential for the interpretation of INS data. It is given by the dynamical structure function 
\be
S(\bk,\omega)=
\frac{1}{\pi}\Bigl[
 {\rm Im} \hchi_{AA}(\bk,\omega)+  {\rm Im} \hchi_{BB}(\bk,\omega)
\Bigr]
.
\ee
This may be evaluated as 
\be
\bl
 S(\bk,\omega > 0)
 &
 =\sum_{\kappa=\pm}I_\kappa(\bk)\delta(\omega-\omega_\kappa(\bk));
 \\
 I_+(\bk)
 &=\frac{\sum_{\si =A,B}m_\si^2\De_\si P_\si(\omega^2_+-\omega^2_{\bar{\si}})}
  {\omega_+(\bk)(\omega_+^2(\bk)-\omega_-^2(\bk))};
 \\
 \;\;\;
 I_-(\bk)
 &=\frac{\sum_{\si =A,B}m_\si^2\De_\si P_\si(\omega^2_{\bar{\si}}-\omega^2_-)}
 {\omega_-(\bk)(\omega_+^2(\bk)-\omega_-^2(\bk))},
 \label{eq:Ising-intens}
 \el
 \ee
 with
 \be
 \bl
 &
\omega_+^2(\bk)-\omega_-^2(\bk)=
\\
&\hspace{0.4cm}
2\Bigl[\frac{1}{4}(\De_A^2-\De_B^2)^2+ 4m_A^2m_B^2\De_A\De_B|I_N(\bk)|^2\Bigr]^\fs,
\el
\ee
where $\bar{\si}=B,A$ for $\si=A,B$. Here $I_\kappa(\bk)$ denotes the bare intensity of each mode in the  INS scattering without Bose-, polarization- and  atomic form factors \cite{jensen:91}; {\blue it will be discussed at the end of Sec.~\ref{sec:Isingbogol}. We note that in RPA method and also in bosonic Bogoliubov approach below the exciton modes are sharp. They may develop a finite broadening or lifetime due to intrinsic excition-exciton interactions \cite{bak:75,jensen:91} or by an extrinsic process originating from the coupling to the electron-hole continuum of (e.g. 5d,6s-type) conduction bands as discussed in detail in Ref. \onlinecite{jensen:91}. Away from the soft mode regime the large exciton gap protects them from overdamping by these processes. However close to the temperature $T_m$ of induced order the softening of $\omega_-(\bQ)$ causes a strong increase of damping channels may  lead to a broadening of the mode into a quasielastic line at the ordering wave vector \cite{holden:74}.
It should be noted that the relation between mode softening and transition to induced order is generally more complicated than predicted by RPA approach \cite{jensen:91}.
}

\subsection{Bosonic representation of interacting CEF excitations}
\label{sec:Isingbogol}

An alternative approach to the magnetic exciton problem is provided by a bosonic representation of the Hamiltonian 
and a subsequent application of Bogoliubov technique for diagonalisation \cite{grover:65}. It has the advantage of not only providing 
the dispersion but also the eigenvectors or Bloch states of magnetic exciton modes. On the other hand it
can only be used a temperatures low compared to the CEF splitting. We first apply it for the simple singlet-singlet
system, restricting for simplicity to $1^\text{st}$ neighbor interactions, in order to use it as a guidance for  the more complicated singlet-doublet system.\\

In the restricted $\Gamma_2-\Gamma_1$ space, considering Eq.~(\ref{eq:matel}) we may replace the angular momentum component $J_z$
by sublattice bosonic operators according to
\be
\bl
J_{iA}^z=m_A(a_i^\dg+a_i);
\quad
 J_{iB}^z=m_B(b_i^\dg+b_i),
\el
\ee
where the $a_i, b_i$ and  $a^\dg_i, b^\dg_i$  satisfy the usual bosonic commutation rules. This replacement produces the proper matrix elements but is restricted to low T because of the different commutation rules and statistics \cite{grover:65,cooper:72,jensen:91,thalmeier:94} {\blue (The thermal occupation of a finite set of CEF states is determined by their Boltzmann factors while the mapping to bosons creates an enlarged space with arbitrary number of excited bosons leading to bosonic statistics.)} Introducing Fourier transforms like  $a_\bk=(1/\sqrt{N})\sum_i\exp(i\bk\bR_i)a_i$ etc. and rearranging terms in the $1^\text{st}$ neighbor exchange Hamiltonian in Eq.~(\ref{eq:HIsing}) we arrive at
\be
\hH=\fs\sum_{\bk}\phi^\dg_\bk\hh_\bk\phi_\bk +E_0;\;\;\; \mbox{with}\;\;\; 
\phi_\bk=(a_\bk,b_\bk,a^\dg_{-\bk},b^\dg_{-\bk})^T,
\ee
here $E_0=(N/2)(\De_A+\De_B)$. The components of this four spinor satisfy the bosonic commutation relations $[\phi_n(\bk),\phi^\dg_m(\bk')]=\Sigma_z^{nm}\delta_{\bk\bk'}$ where $\Sigma_z=\tau_z\otimes 1_2=\mathop{\rm diag}(1_2,-1_2)$ is composed of the $2\times 2$ unit $1_2$.
In this representation we can express
\be
\bl
\hh_\bk=
\left(
 \begin{array}{cccc}
\De_A&-\bint_N^*(\bk)&0&-\bint_N^*(\bk) \\
-\bint_N(\bk)& \De_B &-\bint_N(\bk) & 0\\
0&-\bint_N(-\bk)&\De_A&-\bint_N(-\bk)\\
-\bint^*_N(-\bk)&0&-\bint^*_N(-\bk)&\De_B
\end{array}
\right),
\label{eq:hmatIsing}
\el
\ee
where we used $\bint_N(\bk)=(m_Am_B)I_N(\bk)=(zv_s)\gamma_\bk$ which satisfies $\bint_N(-\bk)=\bint_N(\bk)^*$ (Eq.~(\ref{eq:appstruc2})). The magnetic exciton modes may be obtained by a paraunitary Bogoliubov transformation. The dispersions are then obtained as eigenvalues obtained from the secular equation $|\Sigma_z\hh_\bk-\omega 1|=0$ . The solution of this equation leads to the $T=0$ exciton modes
\be
\bl
\omega^2_\pm(\bk)
&=\fs(\De_A^2+\De_B^2)
\\
&\hspace{0.4cm}
\pm\bigl[\frac{1}{4}(\De_A^2-\De_B^2)^2+
4m_A^2m_B^2\De_A\De_B|I_N(\bk)|^2\bigr]^\fs
.
\label{eq:bogoldisp1}
\el
\ee
The above Eq.~(\ref{eq:bogoldisp1}) is identical to the RPA result for zero temperature $(P_A=P_B=1)$ obtained before in Eq.~(\ref{eq:RPAdisp1}). Therefore on the RPA level one may say that temperature enters in the theory just as a parametric change of the effective exchange coupling by modification of the matrix elements to effective ones with the replacement $m_\si^2\rightarrow P_\si(T)m_\si^2$. In the case of equivalent sublattices A,B the above equation reproduces the $T=0$ case of Eq.~(\ref{eq:RPAdisp2}).{\blue The Bloch functions corresponding to magnetic exciton bands are the eigenvectors of $\Sigma_z\bar{h}_\bk$ corresponding to the four eigenvalues $\pm\omega_\pm(\bk)$.}

At this point, to obtain a preliminary impression of the behaviour of magnetic excitons in the honeycomb lattice we discuss the results for the Ising-type model as presented in Fig.~\ref{fig:isingdisp}. In (a,b) the symmetric case $\Delta_A=\Delta_B$ is shown for elevated (a) and low temperature (b). In the former a moderate dispersion due to small thermal population differences $P_{A,B}$ in Eq.~(\ref{eq:RPAdisp0}) or Eq.~(\ref{eq:RPAdisp2}) exists which becomes larger in the low temperature case. The dispersion of modes is controlled by both by intra- ($v_2$) and inter- ($v_s$) sublattice interaction strength while the mode splitting is only due to the latter (for $v_2^\si=v_2$). At the $\bK_\pm-$ zone boundary points, however they become degenerate because $\gamma(\bK_\pm)=0$ (Appendix \ref{sec:appgeometry}). This degeneracy is lifted by introducing inequivalent A,B CEF splittings as demonstrated in (c,d) for two cases with different strength of intra-sublattice coupling $v_2$. A similar removal of degeneracy at $\bK_\pm$ occurs if the splittings are kept equal but the intra-sublattice couplings $v_2^{A,B}$ become inequivalent.
{\blue The intensity of the modes corresponds to the brightness of the dispersion curves in  Fig.~\ref{fig:isingdisp}.
In particular in (c) one can see that the low energy modes have larger intensity (are brighter) the the high energy modes.
This is due to the mode frequencies appearing in the denominator of intensity expressions in Eq.~(\ref{eq:Ising-intens}).

Experimentally the magnetic exciton dispersion curves are determined by INS \cite{rainford:71,buyers:75,houmann:79,clausen:94}. Comparison with theoretically predicted model dispersions as derived here (Eqs.~\ref{eq:RPAdisp0},\ref{eq:xydisp1}) are the most direct way to extract the physical relevant parameters such as CEF splittings and exchange interaction strengths of the singlet ground state honeycomb material investigated.}

The consistent results of two different techniques in this Section encourage us to consider the more involved and richer singlet-doublet xy-type model. It may also be treated within the response function approach by a simple extension (App. \ref{sec:appgeometry}). It has the drawback of giving only the spectral density of the magnetic excitons but not the composition of the eigenmodes
which is important for discussing topological properties relevant in the $xy$-model. Therefore, in this case we employ the 
bosonic technique in the following.

\section{The singlet-doublet xy -type model}
\label{sec:xymod}

{\blue We outline the aim and according procedure  in this section for clarity: First we define the minimal model ingredients. Then we carry out  the transformation of the Hamiltonian to bosonic coordinates up to bilinear terms (Sec.~\ref{sec:xybosonic}) where, as compared to the Ising case, a doubling of the four-component boson fields occurs due to doublet degeneracy. The magnetic exciton energy bands are then obtained for our most general form of the Hamiltionian (Sec.~\ref{sec:xydispgeneral}). It shows the effects of the various  exchange couplings in the Hamiltonian in a transparent form which will be of great value for extracting their physical value from future experiments on singlet ground state honeycomb materials. The bosonic approach also allows to compute the eigenvectors or Bloch states corresponding to the four exciton bands. These are essential inputs to identify their topological character via the Berry curvature and Chern number as carried out in Sec.~\ref{sec:topol}. 

The exciton dispersions for our most general model  are quite involved. Therefore in Sec.~\ref{sec:reduced} we derive approximate mode energies for the weakly dispersive case sufficiently away from the soft mode regime. We show that in this case the band energies are described by weakly dispersive seperate sublattice modes coupled by the nearest neighbor exchange.
It is also important to consider the general solution for exciton bands for simpler cases to isolate the effect of sublattice symmetry breakings and the presence or absence of the various exchange terms, in particular the DM interaction. This will be carried out in Sec.~\ref{sec:special}. 
At the special zone boundary points $\bK_\pm$ the exciton modes are degenerate unless the DM interaction is nonvanishing. The opening of a bulk gap due to the latter is an important issue in the honeycomb model because it provides the energy window for the appearance of topological edge modes, Therefore we discuss the asymptotic form of bulk bands in the vicinity of the  $\bK_\pm$ points to considerable detail in Sed.~\ref{sec:valley}.

\subsection{Bosonic approach to the magnetic exciton bands of the singlet doublet-model}
\label{sec:xybogol}

In contrast to the Ising model we focus here on the Bogolibuov approach to diagonalise
the model Hamiltoninan. The response function formalism can be applied accordingly
and is described in Appendix \ref{sec:xyRPA}. Our aim is to show that due to the degeneracy
of the excited state it allows for the existence of nontrivial topological character of magnetic exciton bands 
and associated appearance of edge modes within the gap of 2D bulk modes.

\subsubsection{Model Hamiltonian and transformation to bosonic coordinates}
\label{sec:xybosonic}
}

The singlet-doublet model for honeycomb magnetic excitons leads to additional possibilities because of
its xy-type exchange structure as enforced by the selection rules of Eq.~(\ref{eq:matel}). 
They show that in this model
two of the total angular momentum operators $J_x, J_y$ have nonzero matrix elements complementary to the previous
singlet-singlet case that involves only $J_z$. Because the centers of $2^\text{nd}$ neighbor bonds are not inversion centers in any case this opens the possibility for asymmetric DM exchange $H_{DM}=\sum_{\bra\bra ij\ket\ket}\nu_{ij}D_J(J_{ix}J_{jy}-J_{iy}J_{jx})$ according to Moriya rules \cite{moriya:60}. Here we defined $D_J=(g_J-1)^2D$ ($g_J= $Land\'e factor) as the original DM spin-exchange constant $D$ projected to the lowest angular momentum multiplet $(J=4)$ considered in this work.
It has to be staggered along each bond direction 
as expressed by $\nu_{ij}=\pm 1$, i.e. $2^\text{nd}$ neighbors $(-\tbde_i,\tbde_i)$ $(i=1-3)$ have DM exchange $(-D^A_J,D^A_J)$ on A sublattice and conversely $(D^B_J,-D^B_J)$ on the B sublattice. (Fig.~\ref{fig:honeycomb}). The total Hamiltonian in the $\Gamma_2-\Gamma_3$ model is then given by
\be
\bl
H=
\;\;
&\sum_{\Gamma\sigma i}E_\Gamma^\sigma|\Gamma_{\sigma i} \ket\bra\Gamma_{\sigma i}|
-\sum_{\bra ij\ket}I_\si(J^x_{iA}J^x_{jB}+J^y_{iA}J^y_{jB})
\\
&
-\sum_{\bra\bra ij\ket\ket\sigma}I_2^\si(J^x_{i\sigma}J^x_{j\sigma}+J^y_{i\sigma}J^y_{j\sigma})
\\
&
+\sum_{\bra\bra ij\ket\ket\sigma}\nu_{ij}D_J^\si(J^x_{i\sigma}J^y_{j\sigma}-J^y_{i\sigma}J^x_{j\sigma}).
\label{eq:Hxy}
\el
\ee
Here we formulated the most general case of the model with $1^\text{st} (\bra ij\ket)$ and $2^\text{nd} (\bra\bra ij\ket\ket)$ neighbour exchange.{\blue We have in mind symmetric and asymmetric (DM) exchange interactions that are mediated by conduction electrons \cite{yamada:19,togawa:23} . Further allowed exchange interactions like Kitaev terms or symmetric terms off-diagonal in momentum components are suppressed here to keep the number of model constants at a minimum and to isolate the effect of the DMI term.} The CEF splittings as well as the three types of interactions are assumed to be sublattice dependent. As in the Ising case this may be caused by a different chemical environment of the two sublattice sites when the bare 2D honeycomb lattice of 4f ions is integrated into a larger 3D structure. We treat this model again by using  the bosonic representation which is now defined by $(J_\pm=J_x\pm iJ_y)$
\be
\bl
J_+^{iA}=\sqrt{2}\tilm_A(a_{i+}^\dg+a_{i-});\;\;\;J_+^{iB}=\sqrt{2}\tilm_B(b_{i+}^\dg+b_{i-});
\\
J_-^{iA}=\sqrt{2}\tilm_A(a_{i-}^\dg+a_{i+});\;\;\;J_-^{iB}=\sqrt{2}\tilm_B(b_{i-}^\dg+b_{i+}).
\label{eq:bosxy}
\el
\ee
We notice that there is an additional degree of freedom $\lambda=\pm$ corresponding to the two doublet components $|\Gamma_3^\la\ket$ represented by the $a^\dag_{i\la}, b^\dag_{i\la}$ creation operators. Only for some special cases this will remain a degeneracy index throughout the Brillouin zone (BZ) for the diagonalised excitonic eigenmodes.

Now again we introduce the Fourier transformed bosonic operators $a_{\bk\la}, b_{\bk\la}$ and conjugates and express the Hamiltonian of Eq.(\ref{eq:Hxy}) 
through them by using  Eq.(\ref{eq:bosxy}). We finally obtain
\bea
\hH=\fs\sum_{\bk\la}\phi^\dg_{\bk\la}\hh_{\bk\la}\phi_{\bk\la} +E_0,
\eea
with 
$\phi_{\bk\la}=(a_{\bk\la},b_{\bk\la},a^\dg_{-\bk\bla},b^\dg_{-\bk\bla})^T
$.
Here we defined $\bla=-\la$ and $E_0=N(\De_A+\De_B)$. Similar to the Ising-type model the four spinor components satisfy bosonic commutation relations
$[\phi_n(\bk\la),\phi^\dg_m(\bk'\la')]=\Sigma_z^{nm}\delta_{\bk\bk'}\delta_{\la\la'}$
{\blue where the $4\times 4$ diagonal matrix is defined  above Eq.~(\ref{eq:hmatIsing}).} In this representation we now have

\begin{widetext}
\bea
\hh_{\bk\la}=
\left(
 \begin{array}{cccc}
\De_A-\bint^A_D(\bk\la)&-\bint_N^*(\bk)&-\bint^A_D(\bk\la)&-\bint_N^*(\bk) \\[0.2cm]
-\bint_N(\bk)& \De_B-\bint^B_D(\bk\la) &-\bint_N(\bk) &-\bint^B_D(\bk\la) \\[0.2cm]
-\bint^A_D(-\bk\bla)&-\bint_N(-\bk)&\De_A-\bint^A_D(-\bk\bla)&-\bint_N(-\bk)\\[0.2cm]
-\bint^*_N(-\bk)&-\bint^B_D(-\bk\bla)&-\bint^*_N(-\bk)&\De_B-\bint^B_D(-\bk\bla)\\[0.2cm]
\end{array}
\right).
\label{eq:hmatxy}
\eea
\end{widetext}

Here the  intra- (D) and inter- (N) sublattice interactions are defined by
\be
\bl
\bint^A_D(\bk\la)=
&\tilm^2_AI^A_D(\bk\la);
\\
I^A_D(\bk\la)=
&
(z_2I^A_2)\gamma_2(\bk)+\la(z_2D^A_J)\tgam_D(\bk)
\\
=&
I_D^A(-\bk\bla)=I_D^B(-\bk\la) ;
\\
\bint^B_D(\bk\la)=
&\tilm^2_BI^B_D(\bk\la);
\\
 I^B_D(\bk\la)
 =&
 (z_2I^B_2)\gamma_2(\bk)-\la(z_2D^B_J)\tgam_D(\bk)
 \\
=&
I_D^B(-\bk\bla)=I_D^A(-\bk\la);
\\
\bint_N(\bk)=&\tilm_A\tilm_BI_N(\bk);\;\;\;  I_N(\bk)=(zI)\gamma(\bk)
.
\label{eq:interaction}
\el
\ee

{\blue
\subsubsection{General case for magnetic exciton dispersion}
\label{sec:xydispgeneral}
}

Again for numerical computation it is convenient to use (now generally five) model parameters 
$v_s=(\tilm_A\tilm_BI)$, $v_2^\si=(\tilm_\si^2I_2^\si)$ and   $v_D^\si=(\tilm_\si^2I_D^\si)$ and likewise 
$| \bar{I}_N(\bk)|=(zv_s)\gamma(\bk)$, $\bar{I}_D^A(\bk\la)=(z_2v_2^A)\gamma_2(\bk)+\la(z_2v_D^A)\tgam_D(\bk)$
and   $\bar{I}_D^B(\bk\la)=(z_2v_2^B)\gamma_2(\bk)-\la(z_2v_D^B)\tgam_D(\bk)$ (see also Appendix~\ref{sec:parameters}).
Note the sign of the DM term changes with sublattice inversion and $\Gamma_3$ degeneracy index which leads
to the symmetry $\bla\tgam_D(-\bk)=\la\tgam_D(\bk)$ which has been used in the construction of the Hamiltonian matrix 
Eq.(\ref{eq:hmatxy}). The excitonic eigenmodes in the present general model are then, similar as in previous section,
obtained by solving  $|\Sigma_z\hh_\bk-\omega 1|=0$. The solution leads to a closed form of their
dispersions  $\omega_\kappa^2(\bk\la) (\kappa=\pm)$,  given a by formally similar expression as Eq.~(\ref{eq:RPAdisp0}) in the zero temperature limit:
\be
\label{eq:xydisp0}
\bl
&\omega_\pm^2(\bk\la)
=
\fs(\omega_A^2(\bk\la)+\omega_B^2(\bk\la))
\pm
\\
&\hspace{0.3cm}
\bigl[\frac{1}{4}(\omega_A^2(\bk\la)-\omega_B^2(\bk\la))^2+4\tilm_A^2\tilm_B^2\De_A\De_B|I_N(\bk)|^2\bigr]^\fs;
\el
\ee
with
\be\no
\omega_\sigma^2(\bk\la)
=
\De_\si[\De_\si-2\tilm_\si^2I^\si_D(\bk\la)].
\ee
It is, however, distinct from the singlet-singlet model in the following aspects. Firstly, in contrast to the latter the singlet-doublet model can realize the presence of a DM interaction in the intra-sublattice part because two components $J_x,J_y$ have nonzero matrix elements $\tilm$ between $\Gamma_2$ and $\Gamma_3$. Secondly due to the excited state $\Gamma_3$ being a doublet ($\lambda=\pm$)
the number of modes doubles to four. They are still degenerate at each \bk-point for zero DM interaction. For nonzero $D^\si_J$ the modes still fulfill
the symmetry relation $\omega^2_\pm(\bk\la)=\omega^2_\pm(-\bk\bla)$. Furthermore the matrix elements $\tilm_\sigma$ are different from those of the singlet-singlet model $(m_\si)$, see below Eq.(\ref{eq:matel}). Similar as in Sec.~\ref{sec:IsingRPA} the above exciton dispersion  $\omega_\kappa^2(\bk\la) (\kappa=\pm)$  can be written more 
explicitly as 
\be
\bl
&
\omega^2_\pm(\bk\la)
=
\\
&\fs(\De_A^2+\De_B^2)- 
[
\tilm_A^2\De_AI^A_D(\bk\la)+\tilm_B^2\De_BI^B_D(\bk\la)
]
\pm
\\
&
\Bigl\{\bigl[\fs(\De_A^2-\De_B^2)-
(\tilm_A^2\De_AI^A_D(\bk\la)-\tilm_B^2\De_BI^B_D(\bk\la))\bigr]^2
\\
&
+4\tilm_A^2\tilm_B^2\De_A\De_B|I_N(\bk)|^2\Bigr\}^\fs
.
\label{eq:xydisp1}
\el
\ee
When the DM interaction is set to zero and we replace $\tilm_\si\rightarrow m_\si$ and the degeneracy in the $\Gamma_3^\pm$ index $\la$ is ignored this becomes equivalent to the general case of the Ising-type singlet singlet model (Eq.~({\ref{eq:RPAdisp0})). The temperature dependence of the dispersions
can be incorporated by reminding (Sec.\ref{sec:Isingbogol}) that it enters in a parametric way by introducing effective matrix elements $\tilm^2_\si\rightarrow\tilm_\si^2\tanh\frac{\De_\si}{2T}(1+f_\si)^{-1}$ where the correction factor with $f_\si=\fs(1-\tanh\frac{\De_\si}{2T})$ is due to the 
twofold degeneracy of $\Gamma_3$ doublet. This may be concluded from the complementary RPA approach for the xy-type model (Appendix \ref{sec:xyRPA}).\\

\begin{figure}
\includegraphics[width=\linewidth]{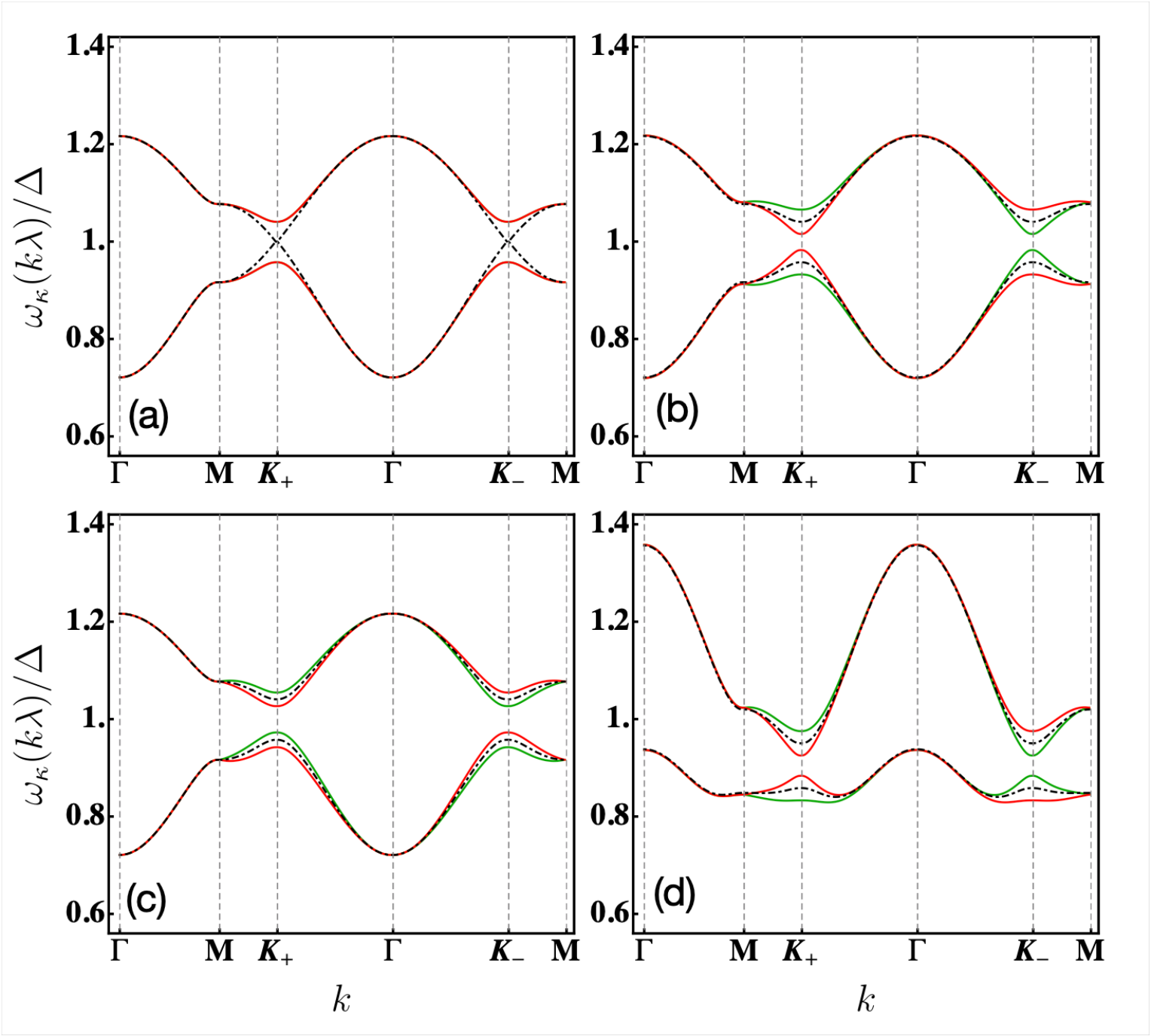}
\caption{Exciton dispersions $\omega_\kappa^\la(\bk)$ ($T=0.1)$ and their typical behaviour of gap formation at \bK$_\pm$ for xy-type model for $\Gamma$MK$_+$$\Gamma$ K$_-$M path in the BZ. Common parameters are $\Delta =1$, $v_s=0.08$.
(a) $v_2=0, \epsilon=0$, dashed line: $v_d=0$; full line: $v_d=0.008$. The DM interaction opens gaps at \bK$_\pm$ but keeps the twofold $\Gamma_3^\pm$ ($\lambda$) degeneracy  throughout the BZ if parameters are identical on sublattices A,B. 
(b) dashed line:  $v_d=0.008, \epsilon=0$; full line:  $v_d=0.008, \epsilon=0.025$ with $\Delta_{A,B}=\Delta(1\pm\epsilon)$). For inequivalent $\Delta_{A,B}$ the $\lambda$ -degeneracy is generally lifted (green: $\la =+$; red: $\la= -$) but prevails along the $\Gamma M$ direction. This is due to the band crossing along $K_+K_-$ segment at $k_y=0$. As a consequence  the band ordering (green/red) is inverted at \bK$_+$ and \bK$_-$.
(c) here $\Delta_A=\Delta_B$ but $v_d^{A,B}=v_d(1\pm\epsilon_d)$ is different with $\epsilon_d=0.35$. This also lifts the \bK$_\pm$ degeneracies but with different sequence of bands. 
(d) This panel corresponds to (b) but now finite (AF) $v_2=-0.03$ included which destroys the approximate reflection symmetry around $\omega=\Delta$.}
\label{fig:xydisp}
\end{figure}

{\blue
\subsubsection{Approximate dispersions from a reduced Hamiltonian}
\label{sec:reduced}
}

The exact expressions for the exciton dispersions of the $4\times 4$ Hamiltonian in Eq.~(\ref{eq:hmatxy}) as given by Eq.~(\ref{eq:xydisp0}) exhibit the redundancy or doubling which is typical for the Bogoliubov technique, i.e. they appear in pairs $(+\omega_\kappa,-\omega_\kappa)$  ( in the RPA response function technique they correspond to poles at positive and negative frequencies). These expressions may be considerably simplified if certain conditions are fulfilled: i) the dispersion width is small compared to the CEF excitation energy $\Delta$ which means that throughout the BZ it is far from soft mode behaviour. This requires $\tilm_\si^2I^\si_D(\bk\la)/\De_\sigma\ll1$.  In this case  $(+\omega_\kappa,-\omega_\kappa)$ pairs are sufficiently apart which means they correspond approximately to the solution of the diagonal $2\times 2$ blocks in $\Sigma_z\hh_{\bk\la}$. This approximation is reasonable if $\Delta_{A,B}$ CEF splittings are not too different. More precisely if we define the various averages
$\Delta_{av}=\fs(\Delta_A+\Delta_B), \bar{\Delta}=(\Delta_A\Delta_B)^\fs, \Delta_m=[\fs(\Delta^2_A+\Delta^2_B)]^\fs$
the conditions $\bar{\Delta}/\Delta_{av}\simeq 1, \Delta_{av}/\Delta_m\simeq 1$ should be respected. For $\Delta_A=\Delta_B$ they hold identically. With these premises the exact dispersions of Eq.~(\ref{eq:xydisp1}) may be approximated by the (positive) exciton energies 
\be
\bl
&\omega^r_\pm(\bk\la)
=
\fs(\omega_{A0}(\bk\la)+\omega_{B0}(\bk\la))
\\
&\hspace{0.3cm}
\pm
\fs\bigl[(\omega_{A0}(\bk\la)-\omega_{B0}(\bk\la))^2+4\tilm_A^2\tilm_B^2|I_N(\bk)|^2\bigr]^\fs,
\\
&
\omega_{\sigma 0}(\bk\la)
=
\Delta_\sigma-\tilm_\si^2I_D^\sigma(\bk\la).
\label{xysimple}
\el
\ee
It can be seen easily that these modes correspond directly to the eigenvalues of the reduced $2\times 2$ Hamiltonian
\bea
\bl
\hh^r_{\bk\la}=
\left(
 \begin{array}{cc}
\De_A-\bint^A_D(\bk\la)&-\bint_N^*(\bk)\\[0.2cm]
-\bint_N(\bk)& \De_B-\bint^B_D(\bk\la)
\end{array}
\right),
\label{eq:redxy}
\el
\eea
which corresponds only to the diagonal blocks in  the $4\times 4$ Hamiltonian Eq.~(\ref{eq:hmatxy}). Effectively the non-diagonal blocks in $\Sigma_z\hh_{\bk\la}$ have the effect of coupling the positive and negative frequency solutions 
$\pm\omega^r_\kappa(\bk\la)$ $(\kappa=\pm)$ of the two diagonal blocks and produce the exact solutions $\pm\omega_\kappa(\bk\la)$ of Eq.~(\ref{eq:xydisp0}) or Eq.~(\ref{eq:xydisp1}). {\blue The approximate treatment of this section provides a convenient starting point for calculating the topological boundary modes in continuum approximation as carried out ins Sec,~\ref{sec:continuum}.}\\

\subsection{Special cases of the singlet-doublet model}
\label{sec:special}

Now we return to the exact and general  dispersion model Eqs.~(\ref{eq:xydisp0},\ref{eq:xydisp1}). We will discuss a few interesting special cases which have either less coupling terms and/or more
sublattice equivalences of model parameters.

{\blue
\subsubsection{First special case}
\label{sec:firstspecial}
}

Here we assume the absence of symmetric $2^\text{nd}$ neighbor exchange and sublattice equivalence of DM terms: $I^\si_2=0, I_D^\si=I_D$.\\

In this case Eq.~(\ref{eq:xydisp1}) reduces to the simpler form
\be
\bl
&\omega^2_\pm(\bk\la)
=\fs(\De_A^2+\De_B^2)-\la(z_2v_D)(\De_A-\De_B)\tgam_D(\bk)
\\
&\pm
\Big\{\frac{1}{4}(\De_A+\De_B)^2[(\De_A-\De_B)-2\la(z_2v_D)\tgam_D(\bk)]^2
\\
&\;\;
+4\De_A\De_B(zv_s)^2|\gamma(\bk)|^2
\Big\}^\fs,
\label{xyspecial1}
\el
\ee
where we introduced abbreviations $v_s=\tilm^2I$ and $v_D=\tilm^2D_J$. This form gives convenient access to the mode dispersions around the inequivalent zone boundary points $\bK_\pm$. The essential part is the `mass term' (first term in curly brackets) given by 
\be
\bl
&
M(\bK_\pm,\la)
=
\\
&\hspace{0.4cm}\fs(\De_A+\De_B)[(\De_A-\De_B)-2\la(z_2v_D)\tgam_D(\bK_\pm)],
\el
\ee
which may be both positive or negative depending on conditions and valley position $\bK_\pm$ (Secs.~\ref{sec:special},\ref{sec:topol}). The above equation shows that in general the $\la$-degeneracy resulting from $\Gamma_3^\pm$ doublet is lifted if firstly, the CEF splittings are inequivalent
and secondly, the DM term is nonzero. This becomes also clear from the next special case:

{\blue
\subsubsection{Second special case}
\label{sec:secondspecial}
}

Here, in addition to the first case we assume the  equivalence $\De_A=\De_B=\De$ :\\
Then we obtain the further simplified dispersion form
\be
\bl
\omega^2_\pm(\bk)
=&\De\bigl\{\De
\pm2[(z_2v_D)^2\tgam_D(\bk)^2+(zv_s)^2|\gamma(\bk)|^2]^\fs\bigr\}
.
\label{xyspecial2}
\el
\ee
Due to the equivalent CEF splittings the dispersions now retain the twofold degeneracy $(\lambda=\pm)$  throughout the BZ, therefore this index has been suppressed. As a result only two dispersion curves ($\kappa=\pm$ due to two sublattices) are present. We also give the simplified dispersion of the reduced model from Eq.~(\ref{xysimple}) for the same special case:
\be
\bl
\omega^r_\pm(\bk)
=\De
\pm[(z_2v_D)^2\tgam_D(\bk)^2+(zv_s)^2|\gamma(\bk)|^2]^\fs
,
\el
\ee
It is obviously the approximation to Eq.~(\ref{xyspecial2}) for moderate dispersion $(v_s,v_D \ll \Delta)$ far from the
soft mode regime.\\

\subsection{Expansions of magnetic exciton dispersion around $\bK_\pm$ valleys}
\label{sec:valley}

It is important to understand the behaviour of exciton bands around the inequivalent valley points $\bK_\pm$ because they influence their topological character. It is largely determined by the expansion of structure functions in Appendix \ref{sec:appstruc}.

{\blue
\subsubsection{General case}
\label{sec:expand-general}
}
For the most general case of parameter sets in Eq.~(\ref{eq:xydisp1}) we obtain the following result  (now $\kappa=\pm$ and $\lambda=\pm$ for the two mode pairs and  $\bK_\pm$ referring now to the two boundary points.):
\be
\bl
\omega^{\kappa 2}_{D}(\bK_\pm,\la,\hat{q})
=
& ~\omega^{2}_{D0}(\bK_\pm,\la)
\\
&+\kappa\{M(\bK_\pm,\la)^2+3\pi^2\De_A\De_B(v_s\hat{q})^2\bigr\}^\fs,
\label{eq:xyexp0}
\el
\ee
where we use the scaled momentum $\hat{q}=(q_x^2+q_y^2)^\fs/(\pi/a)$ with respect to the $\bK_\pm$ Dirac points, i.e. $\bk=\bK_\pm+\bq$. The generally distinct energies of the latter
are given by $(v_\si=\tilm_\si^2I_2^\si, v^\si_D=\tilm_\si^2D^\si_J$ and $\si=A,B)$ 
\be
\bl
\omega^2_{D0}(\bK_\pm,\la)=
&
\fs(\De_A^2+\De_B^2)+ 3(\De_Av_A+\De_Bv_B)
\\
&
\pm\la\sqrt{3}(\De_Av^A_D-\De_Bv^B_D),\quad
\label{eq:DP}
\el
\ee
and depend on valley $(\pm)$ and $\Gamma_3$ degeneracy index $\lambda$.
The splitting of bands at $\bK_\pm$ is determined by the mass term of the square root in Eq.~(\ref{eq:xyexp0})
given by
\be
\bl
M(\bK_\pm,\la)
=
&
\fs(\De_A^2-\De_B^2) +3(\De_Av_A-\De_Bv_B)
\\
&\pm\la\sqrt{3}(\De_Av^A_D+\De_Bv^B_D)
,\quad
\el
\ee
The last term leads to different mass values and (generally) splittings at  $\bK_\pm$ due to its different signs.
{\blue The size of the mode splitting $\delta(\bK_\pm)$ at zone boundary points is given by the difference of the mass terms
for $\la=\pm$, i.e.  $\delta(\bK_\pm)=\pm 2\sqrt{3}(\De_Av^A_D+\De_Bv^B_D)$. It is only finite when the DM interaction 
is nonzero and changes sign between $\bK_\pm$. For the equivalent A,B sublattice model then  $\delta(\bK_\pm)=\pm 4\sqrt{3}\Delta v_D$ the splitting provides a direct means to determine the size of the DMI.} This originates in the different signs of the DM structure function  $\tgam_D(\bK_\pm)=\mp\frac{3\sqrt{2}}{z_2}$ (Appendix \ref{sec:appstruc}).
If the mass term vanishes, the exciton bands are all degenerate at $\bK_\pm$ and show a linear dispersion 
around it due to the last term in  Eq.~(\ref{eq:xyexp0}).\\
Obviously interchanging valley $\bK_\pm$ position and simultaneously the $\Gamma_3$ states $\la=\pm$ leaves the Dirac point energy and mass term invariant, i,e, they fulfil the symmetry $\omega_{D0}(\bK_\pm,\la)=\omega_{D0}(\bK_\mp,-\la)$ and  $M(\bK_\pm,\la)=M(\bK_\mp,-\la)$.\\

As in the previous subsection it is again useful to consider the two special cases with reduced parameter set.

{\blue
\subsubsection{First special case}
}

Here only the CEF splittings are
different on A,B. Then we can simplify, defining the average gap by $\De_{av}=\fs(\De_A+\De_B)$, we have
\be
\bl
\omega^2_{D0}(\bK_\pm,\la)
=
&\fs(\De_A^2+\De_B^2) \pm\la\sqrt{3}v_D(\De_A-\De_B),
\\
M(\bK_\pm,\la)
=
&\De_{av}\bigl[(\De_A-\De_B)\pm\la 2\sqrt{3}v_D\bigr]
.
\label{eq:DPsimple}
\el
\ee
The square of the exciton dispersion is then given by
\bea
\omega^{\kappa 2}_{D}(\bK_\pm,\la,\hat{q})
&=&
\omega^{2}_{D0}(\bK_\pm,\la)+
\kappa[M(\bK_\pm,\la)^2+D_0^2\hat{q}^2]^\fs;\nonumber\\
D_0&=&
\sqrt{3}\pi(\De_A\De_B)^\fs v_s.
\label{eq:dispsimple}
\eea
It is instructive to evaluate directly the dispersion 
$\omega^\kappa_{D}(\bK_\pm,\la,\hat{q})$ at small $\hat{q}$
for the case of finite mass term
\be
\bl
\omega^\kappa_{D}(\bK_\pm,\la,\hat{q})
&=\omega^\kappa_{D0}(\bK_\pm,\la)+\kappa\frac{D_0^2}{4|M|\omega_{D0}}\hat{q}^2;
\\[0.2cm]
\omega^\kappa_{D0}(\bK_\pm,\la)&=\omega_{D0}(\bK_\pm,\la)+\kappa\frac{|M|}{2\omega_{D0}}.
\el
\ee
The first term describes the split energies at the Dirac points or valleys $\bK_\pm$ (first of Eq.~(\ref{eq:DPsimple})). For $\Delta_A\neq\Delta_B$ in Eq.~(\ref{eq:DPsimple}) there are four distinct energies  at each $\bK_\pm$ indexed by $(\kappa,\la)$  and four corresponding split parabolic exciton bands around them (Fig.~\ref{fig:xydisp}(b-d)). \\

\begin{figure}
\includegraphics[width=\linewidth]{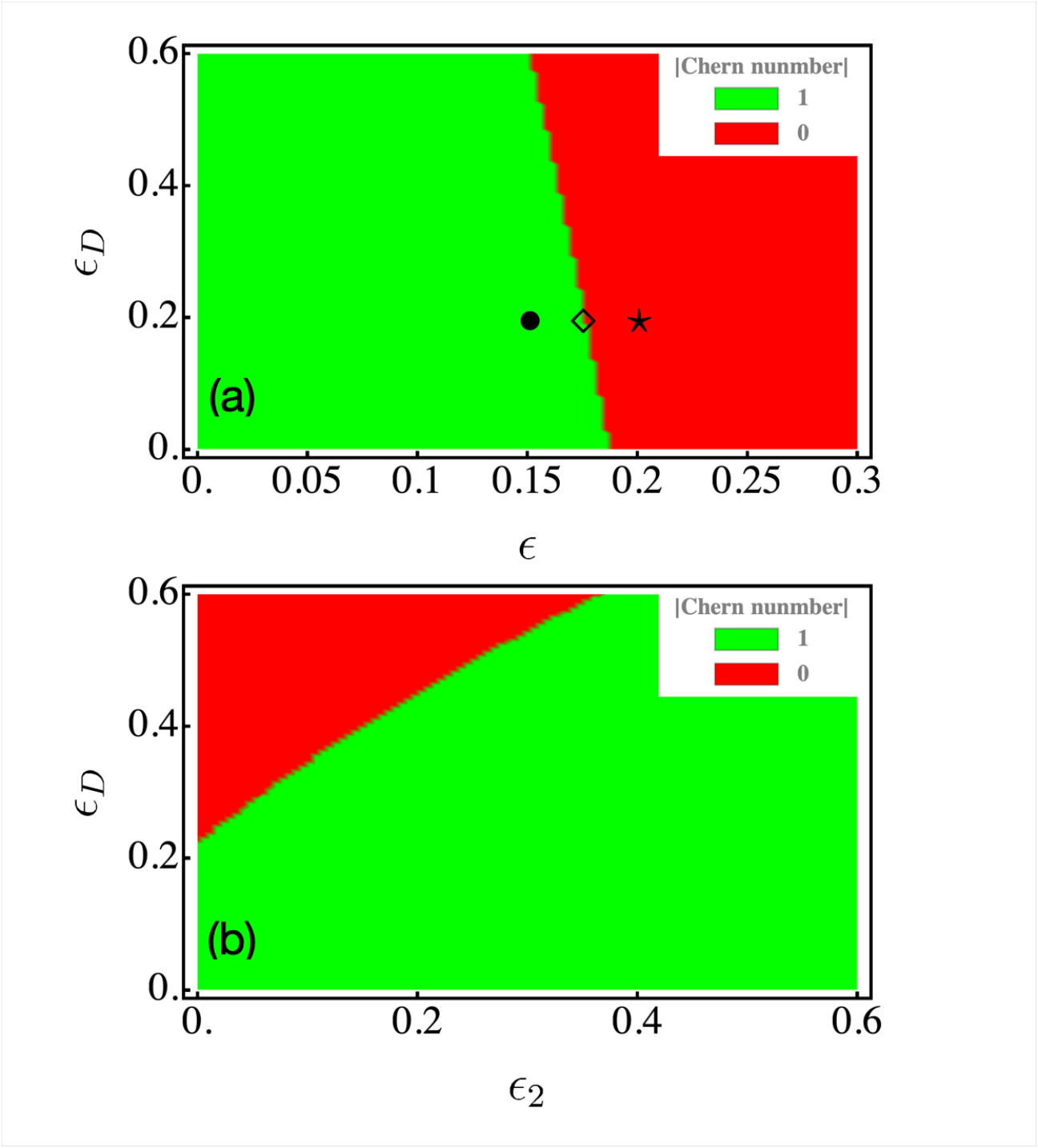}
\caption{
Contour plot of Chern numbers as  functions of
(a) $\epsilon$ in $\Delta_{A,B}=\Delta(1\pm\epsilon)$ and $\epsilon_D$ in $v_D^{A,B}=v_D(1\pm\epsilon_D)$ by setting $\epsilon_2=0$ in  $v_2^{A,B}=v_2(1\pm\epsilon_2)$.
Symbols $\star,\diamond,\bullet$ correspond to values used in Fig.~\ref{fig:berryplot} (a-d), (e-h) and (i-l), respectively.
(b) as functions of $\epsilon_D$, and $\epsilon_2$ by setting $\epsilon=0.175$.
Common parameters are 
$\Delta =1$, 
$v_s=0.08$, 
$v_2=-0.03$, and
$v_D= 0.04$.
The  topological nontrivial  bands are stable in the sublattice equivalent cases, e.g. $(\epsilon,\epsilon_D)=(0,0)$
and  $(\epsilon_2,\epsilon_D)=(0,0)$ .
}
\label{fig:chern}
\end{figure}

{\blue
\subsubsection{Second special case}
}

As in the previous Sec.~\ref{sec:secondspecial} we  assume now in addition equal CEF splittings $\De$ on both A,B sublattice these expressions further  simplify in an obvious manner with $\omega_{D0}=\Delta, M(\bK_\pm,\la)=\pm\la 2\sqrt{3}\De v_D$ and $D_0=\sqrt{3}\pi\De v_s$ which results in two degenerate $(\la=\pm$) pairs of modes.
If we turn off the DM interaction $(v_D=0)$ the mass term vanishes and we have to go back to Eq.~(\ref{eq:dispsimple}) which then leads to
\bea
\omega^\kappa_{D}(\hat{q})=\Delta+\kappa\frac{\sqrt{3}}{2}\pi v_s|\hat{q}|,
\label{eq:dispzero}
\eea
which describes two Dirac half cone $(\kappa=\pm)$ exciton dispersions centered around the CEF excitation energy $\Delta$ which are identical for $\bK_\pm$ and retain the  twofold degeneracy with respect to $\Gamma_3$ index $\la$.\\

\section{Topological properties of magnetic exciton modes}
\label{sec:topol}

Like any kind dispersive modes, in particular magnons in the ferromagnetic honeycomb the paramagnetic exciton bands
studied here can be characterized according to their topological properties. For 2D systems the relevant quantities 
to investigate for this purpose are the Berry curvature and the associated Chern number topological invariant.\\

\subsection{Berry curvature and Chern numbers}
\label{sec:Berry}

The topological character of magnetic exciton bands is determined by Berry curvature obtained from the effective Hamitionian matrix $\tilh(\bk\la)=\Sigma_z\hh(\bk\la)$ (Eq.~(\ref{eq:hmatxy})) which has, for each $\lambda=\pm$ two positive $\omega_\kappa(\bk,\la)$ $(\kappa=\pm,\tau=+)$ and two negative $-\omega_\kappa(\bk,\la)$ $(\kappa=\pm,\tau=-)$ eigenvalues (from Eq.~(\ref{eq:xydisp1})).  The latter are a result of the doubling of degrees of freedom in the Bogoliubov method \cite{heinrich:21}. The index $\tau=\pm$ corresponds to the positive or negative set (the sign in front  of $\tau\omega_\kappa(\bk,\la)$). Then we may combine positive and negative solutions to a single index $n=(\kappa,\tau)=1-4$ resulting from sublattice degree of freedom and Bogoliubov doubling. This is done for each $\lambda=\pm$ subspace resulting from the  $\Gamma_3$ CEF degrees of freedom. The index $\lambda$ is suppressed as a dummy index in the following that simply refers to two different sets of bands (which may be completely degenerate in the BZ as discussed before in special cases). Physical relevant excitations are only the positive energy solutions. The negative solutions however do appear in the calculation of the topological quantities.\\

The topological properties of these bands are described by the Berry curvature given by
\bea
\bOm_n(\bk)=\nabla_\bk\times i\bra n(\bk)|\nabla_\bk|n(\bk)\ket,
\eea
where $|n(\bk)\ket$ denote the {\blue eigenvectors or Bloch functions} corresponding to the eigenvalue equation $\tilh(\bk)|n(\bk)\ket=\omega_n(\bk)|n(\bk)\ket$.
This may also be written as $(\omega_n(\bk)>0)$ \cite{park:19}:
\be
\bOm_n(\bk)=i\sum_{m\neq n}\bra m\bk |\Sigma_z\nabla_\bk |n\bk\ket^*\Sigma^{mm}_z
\times
\bra m\bk |\Sigma_z\nabla_\bk |n\bk\ket
.
\label{eq:berrycurv0}
\ee
An alternative expression more useful for numerical computation is  given by  \cite{park:19}
\be
\bOm_n(\bk)=\sum_{m\neq n}
\frac{i\bra n\bk |\nabla_\bk \hh_\bk |m\bk\ket\Sigma^{mm}_z\times\bra m\bk |\nabla_\bk \hh_\bk |n\bk\ket}
{(\omega_n(\bk)-\omega_m(\bk))^2},
\label{eq:berrycurv1}
\ee
where the sum over $m$ runs over eigenstates with positive {\it and} negative energies $\omega_m(\bk)$. Using
the explicit expression of $\hh_k$ and its gradient $\nabla_\bk\hh_\bk$ as well as the eigenvalues and -vectors of $\tilde{k}_\bk=\Sigma_z\hh_\bk$
the Berry curvature $\bOm_n(\bk)$ may be computed numerically from the above expression. For the 2D honeycomb models only 
the $\Omega^z_n(\bk)$ component is nonzero. Explicitly it is given by
\begin{widetext}
\bea
\Omega^z_n(\bk)=\sum_{m\neq n}
\frac{i[\bra n\bk |\hh^x_\bk |m\bk\ket\Sigma^{mm}_z\bra m\bk |\hh^y_\bk |n\bk\ket-
\bra n\bk |\hh^y_\bk |m\bk\ket\Sigma^{mm}_z\bra m\bk | \hh^x_\bk |n\bk\ket]}
{(\omega_n(\bk)-\omega_m(\bk))^2}
.
\label{eq:berrycurv2}
\eea
\end{widetext}
The Chern number
characterizing the topological character of magnetic exciton bands (reintroducing now the $\Gamma_3$ index $\lambda$) is then obtained by $(n=(\kappa,\tau))$
\be
C_n(\la)=\frac{1}{2\pi}\int_{BZ}d\bk\Omega^z_n(\bk,\la).
\label{eq:chern}
\ee
The \bk-dependence of $\hh_\bk$ in Eqs.~(\ref{eq:hmatIsing},\ref{eq:hmatxy}) stems entirely from that of the structure functions. Therefore the gradients $\hh^\alpha_{\bk\la}=\partial\hh_\bk/\partial k_\alpha$ $(\alpha=x,y)$ required in Eq.~(\ref{eq:berrycurv2}) may be computed analytically (Appendix~\ref{sec:appgrad}). Because the eigenvectors  in Eq.~(\ref{eq:berrycurv2}) have to be obtained numerically this is also necessary for  the Berry curvature. It is shown in Fig.~\ref{fig:berryplot} for some typical parameters for the positive  in the irreducible BZ and will be discussed in more detail in Sec.~\ref{sec:discussion}. There are two typical cases to be observed with Berry curvature maximum (or negative minimum) located at the $\bK_\pm$ zone boundary symmetry points, or at three ($C_{3v}$equivalent) off symmetry points. Whether the Chern number (i.e. the integral of the Berry curvature over the irreducible BZ) is zero (topologically trivial) or nonzero integer (topologically nontrivial) exciton bands depends to some extent on the amount of inversion symmetry breaking (difference of $\sigma =A,B$ sublattice parameters $\Delta_\si,v_2^\si, v_D^\si$, as discussed in Sec.~\ref{sec:discussion}. For the sublattice equivalent case when they are all equal the Chern numbers are all $\pm1$ for the four bands and therefore each of them is topologically nontrivial which should entail the existence of gapless 1D excitonic edge states inside the 2D bulk DM gap at $\bK_\pm$. {\blue The symmetric case is conveniently accessible by a continuum approximation, i.e. small momentum approximation around $\bK_\pm$.  This will indeed predict the existence of edge states as we shall show now.}

\begin{figure*}
\includegraphics[width=0.84\linewidth]{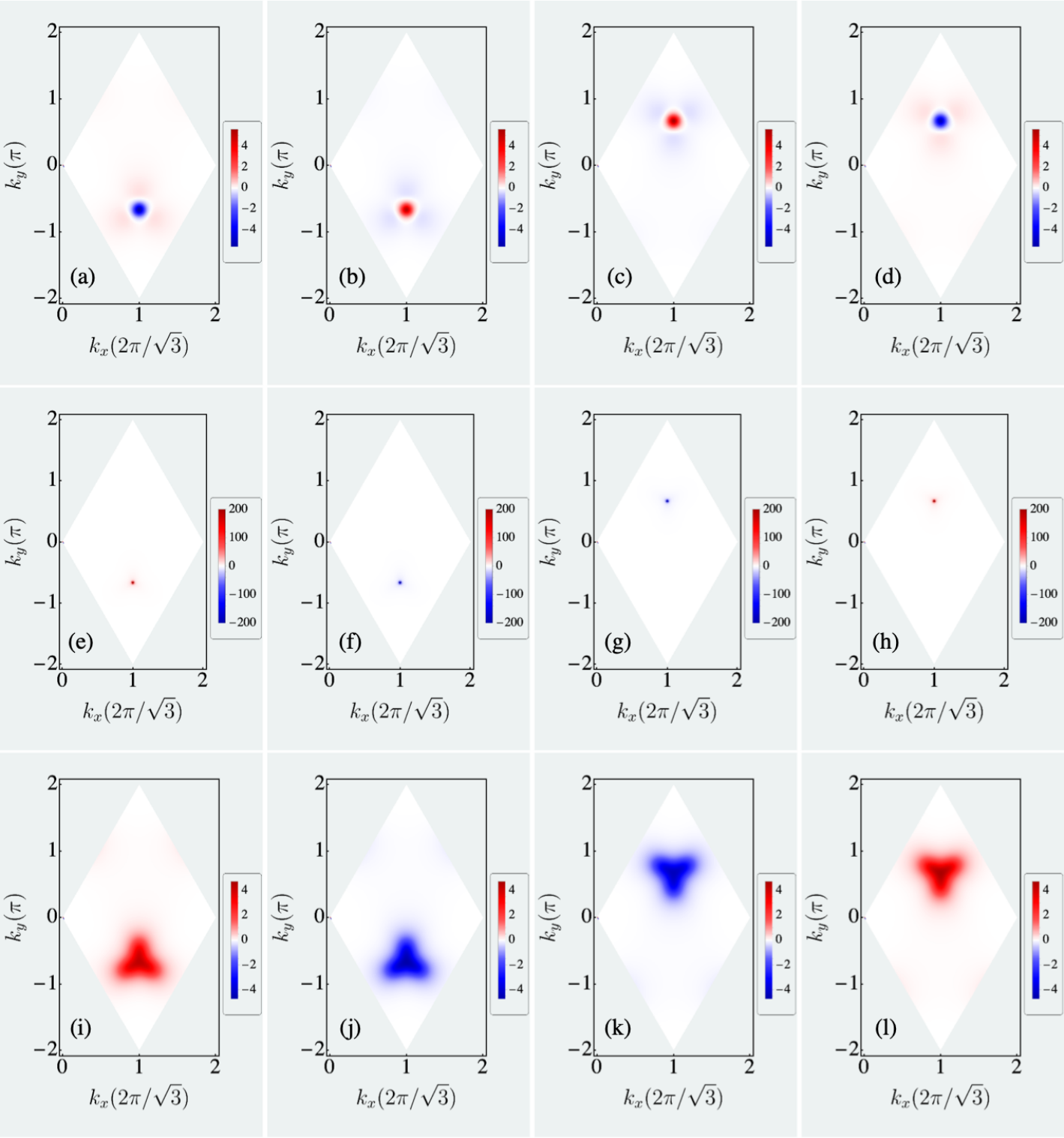}
\caption{
Density plot of Berry curvature in BZ 
corresponding to bands $(\kappa,\lambda)= (+,+), (-,+),(+,-),(-,-)$ from left to right in each row according to 
decreasing energy for each $\lambda$. Parameters are same as  in Fig.~\ref{fig:chern}(a).
First we show two cases according to symbols $(\star,\diamond)$ in Fig.~\ref{fig:chern}(a)
to the right and left of the topological boundary.
(a-d) Chern number $0$ at the $\star$ point $\epsilon=0.2$ and $\epsilon_D=0.2$ in Fig.~\ref{fig:chern}(a);
(e-h) Chern number $\pm 1$  at the $\diamond$ left boundary point $\epsilon=0.175$ and $\epsilon_D=0.2$ in Fig.~\ref{fig:chern}(a).
 Berry curvature has both $\pm$ sign in for each panel of (a-d)  so the integration gives a zero Chern number, however it shows only  positive or negative  values for each panel of  (e-h) and therefore finite Chern number $\pm1$. 
In (i-l) $\epsilon=0.15$ is relatively small
compared to $\epsilon_D$ leading to a shift of the Berry curvature  extrema to three $C_{3v}$ equivalent incommensurate positions
closer to the M-point, however the Chern number still is 
$\pm 1$ corresponding to $\bullet$  in Fig.~\ref{fig:chern}(a).}
\label{fig:berryplot}
\end{figure*}

\subsection{Topological edge modes in continuum approximation}
\label{sec:continuum}

An alternative and direct way to approach the nontrivial topology is provided by the explicit construction
of excitonic magnetic edge states within the 2D bulk gap at $\bK_\pm$ valleys which decay exponentially  into the bulk.
We demonstrate this in the simplified approach mentioned before that neglects the interaction of  $\pm\omega_n(\bk\la)$ modes in the secular equation. This is acceptable as long one is not too close to a soft mode situation. It amounts to considering only the reduced $2\times 2$ Hamiltonian of Eq.~(\ref{eq:redxy}). For the reduced model we apply the continuum approximation around the $\bK_\pm$ by setting $\bk=\bK_\pm +\bq'$ where $\bq'$ is expressed in the rotated Cartesian coordinate systems defined in Appendix~\ref{sec:appstruc}. We first focus on $\bK_+$. The the $q'_{x}$ direction corresponds to zigzag chain direction in real space which we consider as an edge of the semi-infinite honeycomb lattice.
Then we have to replace the perpendicular coordinate according to $q'_{y}\rightarrow -i\partial_{y'}$ in the reduced Hamiltonian above. For the simplified equivalent sublattice case $ii)$ in Sec.~\ref{sec:special} ($v_2=0)$ we obtain
\be
\bl
\hh^r(q'_x\la,y)=
\left(
 \begin{array}{cc}
\De+\la\delta_{D}&(zv_s)\xi(q'_x-\partial_{y'})\\[0.2cm]
(zv_s)\xi(q'_x+\partial_{y'})& \De-\la\delta_{D}
\end{array}
\right),
\el
\ee
where $\xi=\frac{a}{2\sqrt{3}}$ and $\la\delta_{D}=\la3\sqrt{2}v_D$ describes the effect of the DM interaction which importantly has {\it opposite} sign on the two sublattices. As an ansatz wave function for the excitonic edge eigenstate we use $\bw(q'_x,y)=\bw_0e^{iq_x'x'}e^{-\kappa_D y'}$. The corresponding eigenvalue equation $\hh^r(q'_x\la,y)\bw(q'_x,y)=\omega\bw(q'_x,y)$ then leads to the secular equation

\bea
\left|
 \begin{array}{cc}
\De+\la\delta_{D}-\omega&(zv_s)\xi(q'_x+\kappa_D)\\[0.2cm]
(zv_s)\xi(q'_x-\kappa_D)& \De-\la\delta_{D}-\omega
\end{array}
\right| =0,
\label{eq:secedge}
\eea
which has $\lambda$ degenerate solutions 
\be
\bl
\omega_\pm=\Delta\pm\bigl[(\delta_D^2-(zv_s)^2\xi^2\kappa_D^2)+\xi^2(zv_s)^2q^{'2}_x\bigr]^\fs
.
\el
\ee
Choosing $\kappa_D=\frac{|\delta_D|}{zv_s\xi}=\frac{\sqrt{3}}{2}\frac{v_D}{v_s}$ we obtain gapless edge mode dispersions ($\kappa=\pm$)
\be
\bl
\omega_\kappa(q'_x)=\Delta+\kappa(zv_s)\xi|q'_x|=\Delta\pm\frac{\sqrt{3}}{2}\pi v_s|\hat{q}_x|,
\el
\ee
where $\hat{q}_x=q'_x/(\pi/a)$. This describes a 1D Dirac cone of excitonic edge modes emerging from the Dirac point $\omega_{D0}=\Delta$ with momentum oriented along the zigzag chain direction. The calculation is equivalent for the $\bK_-$ value with the replacement $(\delta_D,\kappa_D)\rightarrow (-\delta_D,-\kappa_D)$ in Eq.~(\ref{eq:secedge}) leading to the same dispersion for the edge modes around $\bK_-$. The edge mode dispersion approaches asymptotically
the gapped bulk mode dispersion (for $\Delta_{A,B}=\Delta$)  of Eq.~(\ref{eq:dispsimple}) for $q'_y=0$ and becomes identical to this mode (Eq.~(\ref{eq:dispzero})) when the gap closes $(v_D\rightarrow 0$ and $\kappa_D\rightarrow 0)$.

It is interesting to consider and alternative case of the simplified model without the $2^\text{nd}$ neighbor DM exchange $(v_D=0)$ but instead including the $2^\text{nd}$ neighbor symmetric exchange $(v_2\neq 0)$. In this case the essential difference  from Eq.~(\ref{eq:secedge}) is the lack of  sign change in $\delta_2=3\sqrt{2}v_2$ between the sublattices leading simply to a renormalization of the CEF splitting $\tilde{\Delta}=\Delta+\delta_2$. Therefore the secular equation has no solution for edge states for $q'_x\rightarrow0$ and only bulk states are present.  We conclude that the general structure of the magnetic exciton models discussed here always require a nonzero DM interaction for the existence of topological edge states.

\section{Discussion of numerical results for the xy type model}
\label{sec:discussion}
 
We already discussed the magnetic excitons in the simple Ising type model (Sec.~\ref{sec:Isingmod}) and now focus on the more intricate results of the xy-type model (Sec.~\ref{sec:xymod}).

For a first impression one may restrict to the special models of Sec.~\ref{sec:special}.
the restricted parameter set is then given by CEF splitting energies $\Delta_A, \Delta_B$ and the  sublattice-equivalent interaction energy parameters  $v_s=\tilm^2I$ and $v_D=\tilm^2D_J$ corresponding to $1^\text{st}$ neighbor $(z=3)$ (A-B) symmetric exchange
 and  $2^\text{nd}$ neighbor $(z_2=6)$ (A-A,B-B) DM exchange. The energy unit for these parameters may be chosen
 as the average $\Delta$  and we use the representation $\Delta_{A,B}=\Delta(1\pm\epsilon)$ etc. (Appendix \ref{sec:parameters}).

Some representative dispersion results for these special cases for the xy model are shown in Fig.~\ref{fig:xydisp}.
In (a,b) we also set intra-sublattice $v_2=0$  therefore the splitting of modes caused by inter-sublattice interaction $v_s$ is nearly symmetric around $\Delta$. In (a) when $\epsilon =0$ the upper and lower modes $(\kappa =\pm)$ inherit the twofold degeneracy with respect to the CEF $\Gamma_3$ index $\lambda=\pm$. If the DM interaction vanishes $(v_d=0)$ the two pairs of mode are fully degenerate at $\bK_\pm$ zone boundary points (dashed lines) but for non-vanishing $v_d$ the degeneracy is lifted and a gap appears. The gap persists in the case of inequivalent CEF splittings $(\epsilon \neq 0)$ (b). Now the fourfold degeneracy is completely removed because $\lambda=\pm$ modes are no longer degenerate. This is also true when a finite $v_2$ is included which removes the approximate reflection symmetry of lower and upper branches (d). Note the important point that in both cases the ordering of modes $\lambda =\pm$ (corresponding to the coloring green/red) is interchanged at $\bK_\pm$ This is due to the symmetry $\omega_\kappa(\bk,\lambda)=\omega(-\bk,\bar{\lambda})$ and the fact that $\bK_-$ is equivalent to $-\bK_+$. 
For comparison we also show a case where the CEF splittings are equivalent $(\epsilon=0)$  but the DM coupling strengths  $v_D^{A,B}$ are not, again with $v_2=0$ (c).  It looks similar to (b) but the band ordering is changed such that the gaps at $\bK_\pm$ do not depend on $\lambda$ in contrast to (b).

Now we discuss the topological properties of the magnetic exciton bands. The crucial role there is played by the DM interaction which opens the necessary gap at $\bK_\pm$ for nontrivial topology (nonzero Chern number).
 In the inversion symmetric case with all A,B sublattice parameters equivalent the Chern number is always nonzero
 in the $(v_s,v_D)$ plane as shown in Fig.~\ref{fig:chern}. This agrees with the fact that in the inversion symmetric case the continuum approximation shows the existence of zone boundary modes as shown in the previous section. The introduction of A,B sublattice asymmetry e.g. by assuming different CEF splittings $\Delta_{A,B}=\Delta(1\pm\epsilon)$ 
can destroy the topological state leading to vanishing Chern number as is shown in the example denoted by $\star$ in Fig.~\ref{fig:chern} which shows that the inequivalence of $\Delta_{A,B}$ should stay below a threshold to achieve topologically nontrivial bands with $C=\pm1$. 

To obtain an intuition how the vanishing and non-zero Chern numbers are obtained we also plot the Berry curvature $\Omega_n^z(\bk,\la)$ in the irreducible wedge of the BZ for the different sets of (positive energy) bands $\omega_n(\bk,\la)$ with $n=(\kappa,\tau=+)$ leading to four panels in each row corresponding to all four choices of $(\lambda=\pm,\kappa=\pm)$. We show these four panels for three cases corresponding to the trivial (a-d)  $(C_n(\la)=0)$ and nontrivial (e-h, i-l) $(C_n(\la)=\pm 1)$ regions of Fig.~\ref{fig:chern} marked by symbols $\star,\diamond,\bullet$, respectively.
According to Eq.~(\ref{eq:berrycurv2}) the extremum of Berry curvature occurs close to the points where the exciton band gap is smallest. This naturally happens at $\bK_\pm$ unless the splitting is dominated by the DM interaction as discussed below.
From the dispersion plots in Fig.~\ref{fig:xydisp} it is seen that for a given $\lambda$ the gaps at $\bK_\pm$ are unequal with an inverted order for the opposite $\lambda$. This means the main extremum is situated either on $\bK_-$ or $\bK_+$ for a given $\lambda$. In the trivial case (Fig.~\ref{fig:berryplot}(a-d)) the (absolute) large Berry curvature values at the extrema are compensated by opposite sign values in the surrounding in the irreducible sector integrating to zero Chern number. In the nontrivial case  (Fig.~\ref{fig:berryplot}(e-h)) the sign is the same everywhere and the integration leads to Chern numbers $\pm1$.
Depending on parameters, in particular when DM interaction $v_D$ is large the minimum gap may shift from $\bK_\pm$ to other ($C_{3v}$ equivalent) incommensurate positions closer to the M point in the irreducible BZ sector. Such a case is presented in the Berry curvature plot of  Fig.~\ref{fig:berryplot}(i-l). However the Chern number is still C=$\pm 1$ since one stays in the nontrivial regime of Fig.~\ref{fig:chern}.

Finally we comment on the absence of a thermal Hall effect in the present paramagnetic case. The thermal Hall effect has been proposed and investigated many times \cite{mook:14,kondo:15,owerre:16b,aguilera:20,fujiwara:22,czajka:23} for the FM ordered honeycomb lattice. In this case time reversal symmetry is broken and an intrinsic nonzero thermal Hall current carried by the topological magnonic edge states may appear.
It vanishes however on the antiferromagnetic honeycomb lattice \cite{cheng:16} due to the twofold degeneracy of magnon modes caused by a symmetry operation consisting ot the product of time reversal and inversion \cite{zyuzin:16}). The situation is similar here in the equivalent sublattice model due to the $\lambda$ degeneracy of magnetic excitons. But even in the asymmetric case when all modes are split we have the symmetry $\Omega_z(\bk,\la)= -\Omega_z(-\bk,\bar{\la})$ which can be seen from Fig~\ref{fig:berryplot}. Since the thermal Hall conductivity involves a summation over $\bk,\la$ it will vanish also for the most general case of exciton bands which is consistent with the paramagnetic state.

\section{Summary and conclusion}
\label{sec:conclusion}

In this work we have developed a comprehensive theory of paramagnetic excitons on the honeycomb lattice originating from the localized CEF excitations of f-electron elements  on the two sublattice sites. We assumed a general case where the inversion symmetry may be broken due to different chemical environment of the sublattices. We focused on a model {\it without magnetic order} which may be realized for integer $J$ lanthanide ions like Pr, Tm or U where  the CEF ground state can be  a nonmagnetic singlet. Specifically we treated the $J=4$ based case of an Ising-type singlet singlet model and an xy-type singlet-doublet model allowed by the $C_{3v} $ site symmetry and with CEF splitting energies $\Delta_{A,B}$. The effective inter-site interactions comprise symmetric intra-and inter- sublattice exchange in both models as as well as a new DM type asymmetric exchange for xy-type model allowed by lack of an inversion center on $2^\text{nd}$ neighbor A-A , B-B bonds. These interactions lead to dispersive magnetic excitons in the paramagnetic state with characteristic properties enforced by the underlying honeycomb symmetry. The dispersion increases with decreasing temperature due to the thermal population effect of CEF levels. {\blue We have treated our general model using two alternative techniques, RPA response function method and Bogoliubov bosonic approach and showed that they lead to equivalent results. The latter approach is the suitable one for discussing topological properties of magnetic excitons.}

In the Ising-type case there are two modes which are split by the A,B inter-sublattice  exchange. If inversion symmetry is present the honeycomb structure enforces the degeneracy of these modes at the zone boundary $\bK_\pm$ points. This degeneracy is lifted if the two sublattices become inequivalent (e.g. have different splittings $\Delta_{A,B}$). For sufficiently strong exchange interactions one mode may turn into a precursor soft mode for an induced magnetic order ot the spiral type. The Ising-type model cannot support a DM asymmetric exchange  and therefore its magnetic excitons are  topologically trivial.\\

This changes in the xy-type singlet-doublet model which supports the DM exchange term. The Fourier transform of the asymmetric exchange is non-vanishing at the $\bK_\pm$ points. Due to the doublet degeneracy there are now generally four modes present. The symmetric intersite exchange splits them only into two pairs if A,B sublattices are still equivalent, however even in this case the gap caused by the DM term is preserved at $\bK_\pm$. The remaining pair degeneracy  is lifted throughout the BZ for sublattice-inequivalent CEF splitting or exchange, except along the $\Gamma M$ symmetry direction.

{\blue In the xy-type model a nonzero DM exchange term exists which has not been considered before in the context of  paramagnetic excitons.} It may support  topologically nontrivial magnetic exciton bands even though there is {\it no magnetic order} present.
This distinguishes the present model  from all previous magnetic honeycomb models investigated \cite{kondo:15} which all use (anti-)ferromagnetic order
as precondition to obtain topological magnon states. We have shown that  indeed the nonzero Chern numbers of topological paramagnetic excitons are
stable over a wide range of parameter space, in particular for all parameters in the A,B sublattice equivalent case. The peculiar structure of the underlying Berry curvature in the irreducible BZ sector has been mapped out. Furthermore we have shown within a continuum approximation for the sublattice-symmetric case that magnetic exciton edge modes inside the 2D bulk magnetic exciton gap {\blue caused by DMI} at $\bK_\pm$ exist and their decay length is governed by the ratio of asymmetric DM exchange to symmetric inter-sublattice exchange. This suggests to extend the present analysis and perform  an investigation of edge states of the xy-type magnetic exciton model within a numerical diagonalization approach for various edge and stripe geometries of the honeycomb lattice.
Because of the paramagnetic state time reversal symmetry is not broken and as a consequence these edge modes do not support a thermal Hall effect as another distinction to the magnon topological excitations in the magnetically ordered honeycomb lattice. However it is possible that, as in the magnetically ordered honeycomb models a finite temperature (pseudo-spin) Nernst effect \cite{cheng:16,kim:16,zyuzin:16} may exist in the paramagnetic exciton case which should be investigated based on the analysis in this work. {\blue Furthermore conduction electrons can easily couple to the gapless edge modes. This will modify their spectral properties which may be accessible by STM investigations.}

\begin{acknowledgments}
A.A. greatfully acknowledges Ali G. Moghaddam for useful discussions.\\
\end{acknowledgments}

\appendix

\section{CEF potential with C$_\text{3v}$ symmetry, levels and eigenstates}
\label{sec:appCEF}

Here we discuss to some detail the $J=4$ CEF states for the less common $C_{3v}$ symmetry of the crystalline electric field potential on the honeycomb lattice {\blue because they are , to our knowledge, not easily available in the literature.} The corresponding CEF Hamiltonian is given in terms of Stevens operators $O_n^m(\bJ)$ of the ground state J-multiplet which are polynomials of $(n-m)^\text{th}$ order in $J_z$ and $m^\text{th}$ order in $J_\pm$ {\blue according to Refs.~\onlinecite{hutchings:64,lea:62}. Its structure is determined by the symmetry alone but contains six independent CEF potential parameters $B_n^m$. Formally they may be given in terms of a point charge model simply representing the neighbouring ligands of the f-electron site by Coulomb potentials. The associated charges of the ligands are effective ones screened by the intervening outer-shell (e.g. 5d, 6s) electrons of f-elements \cite{hutchings:64}. In practice the $B_n^m$ have to be determined from adjustment to experimental quantities like low-temperature specific heat, susceptibility in the whole temperature range and spectroscopic results from INS or Raman scattering. The $C_{3v}$ CEF Hamiltonian is given by} 
\begin{eqnarray}
    H_\text{CEF}
    &=&
    B_2^0O_2^0+B_4^0O_4^0+B_6^0O_6^0
    \nonumber\\
    &\phantom{=}&
    +B_4^3O_4^3+B_6^3O_6^3+B_6^6O_6^6.
\label{eq:CEFpot}
\end{eqnarray}
It may be represented as a $(2J+1)\times(2J+1)$ matrix in the space spanned by free ion states $|J,M\ket$ ($|M| \leq J$). If we rearrange the natural sequence (decreasing $M$) of $|J,M\ket$ states suitably $H_\text{CEF}$ can be written in block-diagonal form according to
\begin{widetext}
\begin{equation}
	H_\text{CEF}=
	\left(
\begin{array}{r|ccccccccc}
  & 3 & 0 & -3 & 4 & 1 & -2 & 2 & -1 & -4 \\
\hline
 3 & d_3 & m_{30} & m_{33} & 0 & 0 & 0 & 0 & 0 & 0 \\
 0 & m_{30} & d_0 & -m_{30} & 0 & 0 & 0 & 0 & 0 & 0 \\
 -3 & m_{33} & -m_{30} & d_3 & 0 & 0 & 0 & 0 & 0 & 0 \\
 4 & 0 & 0 & 0 & d_4 & m_{41} & m_{42} & 0 & 0 & 0 \\
 1 & 0 & 0 & 0 & m_{41} & d_1 & -m_{21} & 0 & 0 & 0 \\
 -2 & 0 & 0 & 0 & m_{42} & -m_{21} & d_2 & 0 & 0 & 0 \\
 2 & 0 & 0 & 0 & 0 & 0 & 0 & d_2 & m_{21} & m_{42} \\
 -1 & 0 & 0 & 0 & 0 & 0 & 0 & m_{21} & d_1 & -m_{41} \\
 -4 & 0 & 0 & 0 & 0 & 0 & 0 & m_{42} & -m_{41} & d_4 \\
\end{array}
\right),
\label{eqn:app:hcef}
\end{equation}
\end{widetext}
where the first row and column denote the free ion $M$ value. In terms of the CEF parameters $B_n^m$ the matrix entries are given by
\begin{equation}
\begin{aligned}
	d_4&=28 \left[B_2^0+30 \left(B_4^0+6 B_6^0\right)\right],
	\\
	d_3&=7 \left[B_2^0-180 \left(B_4^0+17 B_6^0\right)\right],
	\\
	d_2&=-8 B_2^0-660 \left(B_4^0-42 B_6^0\right),
	\\
	d_1&=-17 B_2^0+180 \left(3 B_4^0+7 B_6^0\right),
	\\
	d_0&=-20 \left(B_2^0-54 B_4^0+1260 B_6^0\right),
	\\
	m_{41}&=15 \sqrt{14} \left(B_4^3+24 B_6^3\right),
	\\
	m_{42}&=720 \sqrt{7} B_6^6,
	\\
	m_{30}&=9 \sqrt{35} \left(B_4^3-20 B_6^3\right),
	\\
	m_{33}&=2520 B_6^6,
	\\
	m_{21}&=15 \sqrt{2} \left(B_4^3-42 B_6^3\right).
\end{aligned}
\label{eqn:pr:bnm}
\end{equation}
For the eigenvalues and eigenvectors of the three singlets $(\Gamma_{1a,b},\Gamma_2)$ we obtain
\begin{equation}
\begin{aligned}
	E_{1a}&=
	\frac{1}{2} \left(\beta -\sqrt{8 \gamma ^2+\delta ^2}\right),
	\\
	E_{1b}&=
	\frac{1}{2} \left(\beta +\sqrt{8 \gamma ^2+\delta^2}\right),
	\\
	E_2&=
	\alpha .	
\end{aligned}
\label{eqn:eigenvalue}
\end{equation}
where we defined 
\begin{equation}
\begin{aligned}
	\alpha&=d_3+m_{33},
	\\
	\beta&=d_0+d_3-m_{33},
	\\
	\gamma&=m_{30},
	\\
	\delta&=d_0-(d_3-m_{33}).
\end{aligned}
\label{eqn:pr:abc}
\end{equation}
{\blue According to these expressions the singlet-singlet splitting of the Ising-type $\Gamma_{1a,b}$ model 
of Sec.~\ref{sec:Isingmod} is given by $\Delta=\alpha-\frac{1}{2} \left(\beta \mp\sqrt{8 \gamma ^2+\delta ^2}\right)$
and depends, via Eqs.~(\ref{eqn:pr:bnm},\ref{eqn:pr:abc}) on all six $B_n^m$ CEF parameters. And a similar situation
holds for the splitting $\Delta$ of the $\Gamma_2-\Gamma_3$ xy-type singlet-doublet system of Sec.~\ref{sec:xymod}.}\\

Furthermore the corresponding singlet eigenfunctions are given by
\begin{equation}
\begin{aligned}
	\left|\Gamma_{1a}\right\rangle
	&=
	\cos\theta\left|4,0\right\rangle
	+\sin\theta\frac1{\sqrt2}\left(|4,3\rangle-|4,-3\rangle\right),
	\\
	\left|\Gamma_{1b}\right\rangle
	&=
	-\sin\theta\left|4,0\right\rangle
	+\cos\theta\frac1{\sqrt2}\left(|4,3\rangle-|4,-3\rangle\right),
	\\
	\left|\Gamma_2\right\rangle
	&=
	\frac1{\sqrt2}\left(|4,3\rangle+|4,-3\rangle\right),
	\\
	\cos\theta
	&=
	\frac1{\sqrt2}\sqrt{1+\frac1{\sqrt{1+t^2}}},
	\\
	\sin\theta
	&=
	\frac1{\sqrt2}\sqrt{1-\frac1{\sqrt{1+t^2}}},
	\\
	t
	&:=
	\tan(2\theta)=
	\frac{2\sqrt2\gamma}\delta,
	\quad
	0\le\theta\le\frac\pi4.
\end{aligned}
\end{equation}
We note that the {\em anti\/}symmetric linear combination of the $|4,\pm3\rangle$ states belongs to the totally symmetric $\Gamma_1$ representation while the symmetric linear combination belongs to $\Gamma_2$.

Because $\Gamma_2$ is determined by symmetry alone the eigenvalues and -vectors of the remaining singlets $\Gamma_{1a,b}$ are obtained as explicit solutions of a quadratic equation. This factorisation of the original $3\times 3$ matrix problem (upper left block in $H_\text{CEF}$) is due to the fact that two entries $(d_3,d_3)$ appear pairwise. However the second and third block (which give the twofold degenerate levels of the three doublets) the equivalent entries $(d_2,d_4)$ are generally different, therefore the eigenvalues and -vectors result from a true cubic equation. It is too tedious and not useful to give their explicit expressions. In the special case when CEF parameters fulfil a constraint such that $d_2=d_4$ the three doublet eigenvalues will also factorize in one isolated value and a pair resulting from a quadratic equation.

Nevertheless it is possible to parameterize the form of the doublet eigenfunctions. From the second and third block of the matrix representation of $H_\text{CEF}$ in Eq.~(\ref{eqn:app:hcef}) we can read off that they correspond to superpositions like
\begin{equation}
    \left|\Gamma_3^\pm\right\rangle
    =
    u|4,\pm4\rangle+v|4,\mp 2\rangle\pm w|4,\pm1\rangle
\end{equation}
with normalized coefficients $u,v,w$ which we interpret as coordinates of a point on the surface of a 3d unit sphere spanned by the $\left|J,\pm M\right\rangle$ states. Orthonormality is ensured by writing the doublets in the form
\be
\bl
	\left|\Gamma_{3a}^\pm\right\rangle
	&=
	\sin\chi\left(\cos\phi\left|4,\pm4\right\rangle
	+\sin\phi\left|4,\mp2\right\rangle\right)
	\\
	&\phantom{=}
	\pm\cos\chi\left|4,\pm1\right\rangle,
	\\
	\left|\Gamma_{3b}^\pm\right\rangle
	&=
	\left(\cos\alpha\cos\chi\cos\phi-\sin\alpha\sin\phi\right)
	\left|4,\pm4\right\rangle
	\\
	&\phantom{=}
	+\left(\cos\alpha\cos\chi\sin\phi+\sin\alpha\cos\phi\right)
	\left|4,\mp2\right\rangle
	\\
	&\phantom{=}
	\mp\cos\alpha\sin\chi\left|4,\pm1\right\rangle,
	\\
	\left|\Gamma_{3c}^\pm\right\rangle
	&=
	\left(-\sin\alpha\cos\chi\cos\phi-\cos\alpha\sin\phi\right)
	\left|4,\pm4\right\rangle
	\\
	&\phantom{=}
	+\left(-\sin\alpha\cos\chi\sin\phi+\cos\alpha\cos\phi\right)
	\left|4,\mp2\right\rangle
	\\
	&\phantom{=}
	\pm\sin\alpha\sin\chi\left|4,\pm1\right\rangle.
\el
\label{eqn:pr:g3a}
\ee
The three independent angles $\chi$, $\phi$, and $\alpha$ are determined by the three roots of the secular equation of the Hamiltonian doublet block submatrix. The coefficients of the $\left|\Gamma_{3x}^+\right\rangle$ states turn out to be nothing else than the columns of the Euler-angle parametrization of the 3d rotation matrix, associating $\alpha_\text{Euler}\to\phi$, $\beta_\text{Euler}\to\chi$, $\gamma_\text{Euler}\to\alpha$, and the columns like $1\to\left|\Gamma_{3b}^+\right\rangle$, $2\to\left|\Gamma_{3c}^+\right\rangle$, $3\to\left|\Gamma_{3a}^+\right\rangle$. This holds equivalently with $\beta_\text{Euler}\to\pi-\chi$ for the $\left|\Gamma_{3x}^-\right\rangle$ states.

\section{Collection of parameters for numerical calculations}
\label{sec:parameters}

We use parameters that absorb the matrix elements $m_\si$ and $\tilm_\si$ of the Ising and $xy$ cases, respectively into the interaction parameters  so that matrix elements do not appear explicitly. This is done by defining the quantities (dimension of energy) $v_s, v^\si_2$ and $v^\si_D$ ($\si=$A,B sublattice), for brevity we also use the notation $v_2^{A,B}=v_2(1\pm\epsilon_2)$ and  $v_D^{A,B}=v_2(1\pm\epsilon_D)$  in the same manner as we
have used  $\Delta^{A,B}=v_2(1\pm\epsilon)$ before. Here $\epsilon, \epsilon_2$ and $\epsilon_D$ characterize the amount of inversion symmetry breaking between the sublattices. There are three (five) possible Ising ($xy$) model parameters given by  (coordination numbers $z=3$, $z_2=6$):\\

{\it Ising-type model:}
\be
\bl
v_s=&(m_Am_BI);
\\
v_2^\si
=&(m_\si^2I_2^\si),
\el
\ee
leading to 
\be
\bl
m_Am_B| I_N(\bk)|=&(zv_s)|\gamma(\bk)|
\\
m_\si^2I_D^\si(\bk)=&(z_2v_2^\si)\gamma_2(\bk).
\el
\ee

 {\it xy-type model:}
 \be
 \bl
v_s=&(\tilm_A\tilm_BI) ;
\\
v_2^\si
=&(\tilm_\si^2I_2^\si);
\\
v_D^\si
=&(\tilm_\si^2I_D^\si),
\el
\ee
leading to
\be
\bl
&
\tilm_A\tilm_B| I_N(\bk)|=|\bar{I}_N(\bk)|
=(zv_s)|\gamma(\bk)|,
\\
&
\tilm_A^2I_D^A(\bk\la)=\bar{I}_D^A(\bk\la)
=(z_2v_2^A)\gamma_2(\bk)+\la(z_2v_D^A)\tgam_D(\bk),
\\
&
\tilm_B^2I_D^B(\bk\la)=\bar{I}_D^B(\bk\la)
=(z_2v_2^B)\gamma_2(\bk)-\la(z_2v_D^B)\tgam_D(\bk)
.
\el
\ee
It is clear that a full consideration of the model in the five-parameter space would be too exhaustive. Therefore
only typical cases will be considered with some sublattice parameters equal and/or some parameters set to zero.\\
In the definition of the Hamiltonians we choose the convention that positive $I,I_2^\si$  corresponds to FM
exchange and negative ones to AF exchange. The same convention applies then to $v_s$ and $v_2^\si$ if we make
the reasonable restriction that $m_A$ and $m_B$ matrix elements have the same sign. The sign of $I_D^\si$ is not essential as the DM interaction alternates from bond to bond and from A to B. A change in sign of $I_D^\si$ or $v_D^\si$  just means a redifinition of $\la\rightarrow -\la$ notation in the exciton bands.

\section{RPA response function approach for the xy-type model}
\label{sec:xyRPA}

In this model the twofold $\Gamma^\pm_3$ excited state degeneracy $(\la=\pm)$ and two sublattices lead in principle to a $4\times 4$ susceptibility matrix, which however is the direct sum of $2\times 2$ matrices so that instead of Eq.(\ref{eq:RPA})
we now have
\be
\bl
\hat{\chi}(\bk,\la,\om)
&
=[1-\hat{I}(\bk\la)\hat{u}({\om})]^{-1}\hat{u}({\om});
\\
 \hat{u}({\om})=
 &
\left(
 \begin{array}{cc}
 u_A(\om)& 0 \\
 0& u_B(\om)
\end{array}
\right);
\\
\hat{I}({\bk\la})=
&
\left(
 \begin{array}{cc}
 I_D^A(\bk\la)& I_N(\bk) \\
 I_N^*(\bk)& I_D^B(\bk\la)
\end{array}
\right)
,
\label{eq:RPAxy}
\el
\ee
where the exchange matrix elements are defined in Appendix~\ref{sec:parameters} above.
The single ion susceptibility (the sum of $xx$ and $yy$ components) is given by
\be
\bl
u_{\sigma}(\om)=\frac{2\tilm^2_\sigma\De_\sigma P_\si(T)}
{\De^2_\sigma-(\om)^2}.
\el
\ee
Now the thermal population factor for the singlet-doublet case  is 
 $P_\si(T)=\tanh\frac{\De_\si}{2T}(1+f_\si)^{-1}$ where $f_\si=\fs(1-\tanh\frac{\De_\si}{2T})$. The poles of the 
 dynamical susceptibility associated with magnetic exciton modes may then be obtained in a completely analogous
 way to the Ising model case, except for the additional mode index $\lambda$ resulting from the $\Gamma_3$ degeneracy:
\be
\bl
\omega_\pm^2(\bk\la)=
&\fs(\omega_A^2(\bk\la)+\omega_B^2(\bk\la))
\\
&
\pm
\Bigl[
\frac{1}{4}(\omega_A^2(\bk\la)-\omega_B^2(\bk\la))^2
\\
&
+4\tilm_A^2\tilm_B^2\De_A\De_BP_AP_B|I_N(\bk)|^2
\Bigr]^\fs;
\\
\omega_\sigma^2(\bk\la)=\;
&\De_\si[\De_\si-2\tilm_\si^2P_\si I^\si_D(\bk\la)].
\el
\ee
In the zero temperature limit $P_\si \rightarrow 1$ and the above expression is completely equivalent to 
the xy-model exciton dispersions obtained from the Bogoliubov approach (Eq.~(\ref{eq:xydisp0})). Likewise the
spectral function of the magnetic response is given in an obvious generalization as
\be
\bl
S(\bk,\omega)=\frac{1}{\pi}\sum_\la\Bigl(  {\rm Im} \hchi_{AA}(\bk\la,\omega)+  {\rm Im} \hchi_{BB}(\bk\la,\omega)\Bigr).
\el
\ee

\section{Geometric properties of honeycomb lattice and Brillouin zone}
\label{sec:appgeometry}

The honeycomb lattice (Fig.\ref{fig:honeycomb}) has two basis atoms denoted by A,B with a distance d apart (n.n. distance A-B).
The lattice constant is denoted by a (n.n.n. distance A-A or B-B). They are related by $d=a/\sqrt{3}$. We 
generally use the lattice constant $a$ in the direct lattice and $2\pi/a$ in the reciprocal lattice as units.
The three vectors to n.n. sites $\bde_i$ and  and six vectors to n.n.n. sites $\pm\tbde_i$ (i=1-3) are given by
\be
\bl
\bde_1=&\bigl(\frac{\sqrt{3}}{6},\fs)a;\;\;\;\bde_2=\bigl(\frac{\sqrt{3}}{6},-\fs)a;\;\;\;\bde_3=\bigl(-\frac{\sqrt{3}}{3},0)a;
\\
\tbde_1=&\bigl(\frac{\sqrt{3}}{2},\fs)a;\;\;\;\tbde_2=\bigl(-\frac{\sqrt{3}}{2},\fs)a;\;\;\;\tbde_3=\bigl(0,-1)a
.
\el
\ee
As basis vectors of the unit cell and lattice we may use $\bv_1=-\tbde_2, \bv_2=\tbde_1$. The reciprocal lattice vectors $\bG_1,\bG_2$ are then defined 
via $\bv_i\cdot\bG_j=2\pi\delta_{ij}$ $(i,j=1,2)$. Explicitly we have
\be
\bl
\bv_1=&-\tbde_2=\bigl(\frac{\sqrt{3}}{2},-\fs)a;\;\;\;\bv_2=\tbde_1=\bigl(\frac{\sqrt{3}}{2},\fs)a;
\\
\bG_1=&\bigl(\frac{\sqrt{3}}{3},-1)\frac{2\pi}{a};\;\;\;\bG_2=\bigl(\frac{\sqrt{3}}{3},1)\frac{2\pi}{a}
.
\el
\ee
For the direct unit cell volume we have $V_c=|\bv_1\times\bv_2|=\frac{\sqrt{3}}{2}a^2$ and likewise for the reciprocal cell volume
$\Omega_c=|\bG_1\times\bG_2|=\frac{2}{\sqrt{3}}\bigl(\frac{2\pi}{a}\bigr)^2$ which fulfil the relation $V_c\cdot\Omega_c=(2\pi)^2$.
The inequivalent zone boundary vectors $\bK_\pm$ are given by
\be
\bl
\bK_+
=&\frac{1}{3}[\bG_1+2\bG_2] = \bigl(\frac{\sqrt{3}}{3},\frac{1}{3}\bigr)\frac{2\pi}{a},
\\
\bK_-
=&\frac{1}{3}[2\bG_1+\bG_2] =\bigl(\frac{\sqrt{3}}{3},-\frac{1}{3}\bigr)\frac{2\pi}{a}.
\el
\ee

\section{properties of momentum dependent honeycomb structure functions}
\label{sec:appstruc}

The momentum dependence and in particular gap existence of exciton modes at the zone boundary is determined by
the structure functions of the nearest and next-nearest neighbor interactions depicted in Fig.~\ref{fig:honeycomb}.
They are given by
\be
\bl
&
\gamma(\bk)=
\frac{1}{z}\sum_{\bde}\exp(i\bk\cdot\bde);\;\;\;\gamma_2(\bk)=\frac{1}{z_2}\sum_{\tbde}\exp(i\bk\cdot\tbde);
\\
&\gamma^{A,B}_D(\bk)
=
\frac{1}{z_2}\sum_{\tbde}\nu^{A,B}_{\tbde}\exp(i\bk\cdot\tbde)=:\mp i\tgam_D(\bk),
 \label{eq:appstruc1}
\el
\ee
where $\gamma(\bk)$ and $\gamma_2(\bk)$ correspond to the symmetric $1^\text{st} (\bde)$ and $2^\text{nd} (\tbde)$ neighbor exchange,
with coordination numbers $z=3$ and $z_2=6$, respectively, whereas $\tilde{\gamma}_D(\bk)$ is associated with the asymmetric DM exchange with  second neighbors. The first one
is complex with $\gamma(-\bk)=\gamma^*(\bk)$ the second one is real and even $\gamma(-\bk)=\gamma(\bk)$  while the latter is real and odd $\tgam_D(-\bk)=-\tgam_D(\bk)$  under inversion. The latter is due to the staggered nature of the DM iinteraction leading to $\nu_{\tbde}=-\nu_{-\tbde}=\pm 1$ and $\nu^B_{\tbde}=-\nu^A_{\tbde}$. Explicitly we have, from 
 Fig.~\ref{fig:honeycomb}.:
 \be
 \bl
 \gamma(\bk)=
 &\frac{1}{3}\bigl[\exp{i(\fhxc ak_x +\fs ak_y)} 
 \\
 &
 +\exp{i(\fhxc ak_x -\fs ak_y)}+\exp{(-i\fhxb ak_x)}  \bigr],
 \\
 \gamma_2(\bk)
 =
 &\frac{1}{3}\bigl[\cos(\fhxa ak_x +\fs ak_y) 
 \\&
 +\cos(-\fhxa ak_x +\fs ak_y)+\cos ak_y \bigr],
 \\
 \tgam_D(\bk)=
 &\frac{1}{3}\bigl[\sin(\fhxa ak_x +\fs ak_y) +\sin(-\fhxa ak_x +\fs ak_y)
 \\
 &
 -\sin ak_y \bigr]
 .
 \label{eq:appstruc2}
  \el
  \ee
 It is important to know the behaviour of the structure functions around the zone boundary valleys 
 $\bK_\pm=\bigl(\frac{\sqrt{3}}{3},\pm\frac{1}{3})\frac{2\pi}{a}$. We express the momentum by $\bk=\bK_\pm+\bq$ with $|\bq| \ll \frac{\pi}{a}$.
 Then the structure functions in Eq.~(\ref{eq:appstruc2}) may be expanded in terms of \bq~to lowest order. It is more convenient to use
 hexagonal coordinates $\bq'=(q'_x,q'_y)$ instead of the Cartesian $(q_x,q_y)$. The transformations between them, for each $\bK_\pm$ are given
 by
 \be
 \bl
 \bK_+: q'_x
 =
 &\fs(\sqrt{3}q_x+q_y);\;\;\;q'_y=-\fs(q_x-\sqrt{3}q_y);
 \\
 \bK_-: q'_x
 =
 &\fs(\sqrt{3}q_x-q_y);\;\;\;q'_y=\fs(q_x+\sqrt{3}q_y)
 .
 \el
 \ee
 Then the expansion leads to
 \be
 \bl
 \gamma(\bk)=
 &\gamma(\bK_\pm+\bq')=-\frac{a}{2\sqrt{3}}(q'_x\pm iq'_y),
 \\
 |\gamma(\bk)|^2=
 &\frac{a^2}{12}(q_x^2+q_y^2)=\frac{a^2}{12}({q'_x}^2+{q'_y}^2)=\frac{\pi^2}{12}\hat{q}^2,
 \\[0.4cm]
 \gamma_2(\bk)=
 &\gamma_2(\bK_\pm)=-\frac{3}{z_2},
 \\
 \tgam_D(\bk)=
 &\tgam_D(\bK_\pm)=\mp\frac{3\sqrt{2}}{z_2},
 \el
 \ee
 where we defined $\hat{q}=(q_x^2+q_y^2)^\fs/(\pi/a)$.
 The lowest order term in $\gamma(\bk)$ is the term linear in $\bq'$~ because $\gamma(\bK_\pm)=0$. On the other hand $\gamma_2(\bk)$ and $\tgam_D(\bk)$ have finite values at $\bK_\pm$ and no linear terms in $\bq'$. Note that importantly $\tgam_D(\bK_\pm)$ changes sign between the nonequivalent BZ boundary points.
 
 \section{Momentum gradients of structure functions and Hamiltonian}
\label{sec:appgrad}
 
 The Hamiltonian gradients
 $\hh^\alpha_{\bk\la}=\partial\hh_\bk/\partial k_\alpha$ $(\alpha=x,y)$
 appearing in the matrix elements for the Berry curvature $\Omega_n^z(\bk)$ of Eq.~(\ref{eq:berrycurv2}) are entirely determined by those of the structure functions $\gamma^\alpha(\bk)=\partial\gamma(\bk)/\partial k_\al$ and likewise for $\gamma_2(\bk)$ and  $\tgam_D(\bk)$. From Eq.~(\ref{eq:appstruc2})
 we get for first neighbors
 \be
 \bl
 \gamma^x(\bk)=
 &i\bigl(\frac{a\sqrt{3}}{6}\bigr)\frac{1}{3}
 \bigl[\exp{i(\fhxc ak_x +\fs ak_y)} 
  \\
  &
  +\exp{i(\fhxc ak_x -\fs ak_y)}-2\exp{(-i\fhxb ak_x)}  \bigr],
 \\
 \gamma^y(\bk)=
 &i\bigl(\frac{a}{2}\bigr)\frac{1}{3}
 \bigl[\exp{i(\fhxc ak_x +\fs ak_y)} 
   \\
  &
  -\exp{i(\fhxc ak_x -\fs ak_y)}  \bigr],
 \label{eq:deriv1}
 \el
 \ee
 and for second neighbors
 \bea
 \bl
 \gamma^x_2(\bk)=
 &
 \bigl(-\frac{\sqrt{3}a}{2}\bigr)\frac{1}{3}
 \bigl[\sin(\fhxa ak_x +\fs ak_y) 
   \\
  &
  -\sin(-\fhxa ak_x +\fs ak_y) \bigr],
  \\
 \gamma^y_2(\bk)=
 &
 \bigl(-\frac{a}{2}\bigr)\frac{1}{3}
 \bigl[\sin(\fhxa ak_x +\fs ak_y) 
 \\&
 +\sin(-\fhxa ak_x +\fs ak_y) 
  +2\sin(ak_y) \bigr]
  ,
  \\
 \tgam_D^x(\bk)
 =
 &
 \bigl(\frac{\sqrt{3}a}{2}\bigr)\frac{1}{3}
 \bigl[\cos(\fhxa ak_x +\fs ak_y) 
   \\
  &
  -\cos(-\fhxa ak_x +\fs ak_y) \bigr],
 \\
 \tgam_D^y(\bk)=
 &
 \bigl(\frac{a}{2}\bigr)\frac{1}{3}
 \bigl[\cos(\fhxa ak_x +\fs ak_y) 
    \\
  &
  +\cos(-\fhxa ak_x +\fs ak_y) 
  -2\cos(ak_y) \bigr]
  .
 \label{eq;deriv2}
 \no
 \el
 \\
 \eea

 Then, using Eq.~(\ref{eq:hmatxy})  we obtain the Hamiltonian gradients as \\
\be
\bl
\hh^\al_{\bk\la}=
\left(
 \begin{array}{cccc}
-\bint^{A\al}_D(\bk\la)&-\bint_N^{\al*}(\bk)&-\bint^{A\al}_D(\bk\la)&-\bint_N^{\al*}(\bk) \\[0.2cm]
-\bint_N^\al(\bk)& -\bint^{B\al}_D(\bk\la) &-\bint_N^\al(\bk) &-\bint^{B\al}_D(\bk\la) \\[0.2cm]
\bint^{A\al}_D(-\bk\bla)&\bint_N^\al(-\bk)&\bint^{A\al}_D(-\bk\bla)&\bint_N^\al(-\bk)\\[0.2cm]
\bint^{\al*}_N(-\bk)&\bint^{B\al}_D(-\bk\bla)&\bint^{\al*}_N(-\bk)&\bint^{B\al}_D(-\bk\bla)\\[0.2cm]
\end{array}
\right),
\label{eq:hxyderiv}
\el
\ee
and the interaction derivatives are obtained from Eq.~(\ref{eq:interaction}) as
\be
\bl
\bint^{A\al}_D(\bk\la)
=&
\tilm^2_AI^{A\al}_D(\bk\la),
  \\
  I^{A\al}_D(\bk\la)
  =&
  (z_2I^A_2)\gamma^\al_2(\bk)+\la(z_2D^A_J)\tgam^\al_D(\bk)
    \\
  =&
  -I_D^{A\al}(-\bk\bla)=   -I_D^{B\al}(-\bk\la) ,
  \\
\bint^{B\al}_D(\bk\la)
=&
\tilm^2_BI^{B\al}_D(\bk\la),
\\
I^{B\al}_D(\bk\la)=&(z_2I^B_2)\gamma^\al_2(\bk)-\la(z_2D^B_J)\tgam^\al_D(\bk)
\\
=&
-I_D^{B\al}(-\bk\bla)
=-I_D^{A\al}(-\bk\la),
\\
\bint_N^\al(\bk)
=&
\tilm_A\tilm_BI_N^\al(\bk),
\\
 I_N^\al(\bk)
 =&
 (zI)\gamma^\al(\bk)
 .
\label{eq:intderiv}
\el
\ee
\\

\bibliography{References}

\end{document}